\documentclass[journal]{IEEEtran}
\usepackage{amsmath,amssymb,amsfonts,amsthm,mathtools}
\usepackage{graphicx}
\usepackage{xcolor}
\usepackage{cite}
\usepackage{url}
\usepackage{booktabs}
\usepackage{array}
\usepackage{tabularx}
\usepackage{multirow}
\usepackage{textcomp}
\usepackage{stfloats}
\usepackage{verbatim}
\usepackage[caption=false,font=footnotesize]{subfig}
\usepackage{algorithm}
\usepackage{algorithmic}
\usepackage[long]{optidef}
\usepackage{steinmetz}
\usepackage{bigints}
\usepackage{microtype}
\usepackage[hidelinks]{hyperref} 

\begin{document}
\title{Adaptive RSMA-OMA for Resilient MIMO Networks Under Imperfect CSI and SIC}
\author{
Sayanti~Ghosh,~\IEEEmembership{Member,~IEEE}, 
Indrakshi~Dey,~\IEEEmembership{Senior~Member,~IEEE}, 
and Nicola~Marchetti,~\IEEEmembership{Senior~Member,~IEEE}

\thanks{S.~Ghosh and N.~Marchetti are with the Department of Electrical and Electronic Engineering, Trinity College Dublin, Dublin, Ireland (e-mail: saghosh@tcd.ie; nicola.marchetti@tcd.ie).}

\thanks{I.~Dey is with the Walton Institute, South East Technological University, Waterford, Ireland (e-mail: indrakshi.dey@waltoninstitute.ie).}

\thanks{This work was supported in part by the EU MSCA Project ``COALESCE'' under Grant 101130739, by the US--Ireland R\&D Partnership Programme RI-SFI-23/US/3924, and by Science Foundation Ireland under Grant 13/RC/2077\_P2.}
}
\maketitle
\begin{abstract}
This paper addresses the challenge of power control in Rate-Splitting Multiple Access (RSMA) systems for downlink Multi-Input Multi-Output (MIMO) networks under practical impairments such as spatial correlation, imperfect Channel State Information (CSI), and residual Successive Interference Cancellation (SIC) errors. We propose a novel degeneracy-aware framework that adaptively adjusts the power allocation between the common and private streams, ensuring optimal performance despite CSI uncertainty and imperfect SIC. Our approach incorporates a dynamic switching mechanism between RSMA and Orthogonal Multiple Access (OMA) to maintain system feasibility and resilience in the face of these impairments. Extensive analytical and simulation results demonstrate that the proposed framework significantly enhances power efficiency, mitigates outage probability, and improves overall system robustness, making RSMA a viable and efficient solution for modern wireless networks with realistic CSI and SIC conditions.
\end{abstract}
\section{Introduction}
\IEEEPARstart{T}{he} next generation of wireless networks, envisioned under the IMT-2030 framework, aims to achieve order-of-magnitude improvements in spectral efficiency, reliability, scalability, and energy efficiency, while supporting massive connectivity and highly heterogeneous traffic
profiles \cite{IMT2030,ITUR6G}. Beyond traditional mobile broadband, future systems are expected to seamlessly enable immersive extended reality, digital twins, connected robotics, intelligent transportation, and large-scale machine-type communications.
These applications impose stringent requirements, including ultra-high
data rates, sub-millisecond latency, extreme reliability, and enhanced network resilience under dynamic and unpredictable operating conditions. Meeting such ambitious performance targets necessitates fundamentally
new interference management and multi-access (MA) strategies that remain effective under mobility, spatial congestion, infrastructure failures, and channel uncertainty.
The integration of multi-antenna transmission with flexible MA techniques is a key enabler for improving spatial reuse and throughput \cite{Arun2021}. However, conventional orthogonal multiple access (OMA) and non-orthogonal multiple access (NOMA)/space division multiple access (SDMA) schemes are sensitive to channel state information (CSI) accuracy and interference dynamics. In ultra-reliable low-latency communication scenarios, standard fading models limit channel predictability and require high-SNR margins to mitigate unmodeled uncertainties \cite{Swamy2019}. This highlights the limitation of rigid transmission strategies, where imperfect CSI or fluctuating interference cause rate saturation, fairness issues, and reduced robustness.

Rate-Splitting Multiple Access (RSMA) has emerged as a unifying paradigm bridging the gap between decoding and treating interference as noise \cite{Alea2023,Mishra2022,Mao2022}. By splitting user messages into common and private components, and optimizing power allocation and decoding order, RSMA generalizes SDMA and NOMA. It improves robustness to imperfect CSI at transmitter (CSIT), enhances fairness, and boosts spectral efficiency. Literature shows that RSMA improves max-min fairness in multiuser multiple-input-single-output networks with CSIT uncertainty \cite{Joudeh2016}, and achieves gains in unicast–multicast systems \cite{Mao2019}, cloud radio access network \cite{Robert2022,Ahmad2020}, and mixed-criticality networks \cite{Reifert2022}.  With increasing demand for high rates, RSMA offers flexible interference management and better spectral efficiency compared to OMA schemes \cite{Clerckx2019RSMA}.

Beyond spectral efficiency, resilience in next-generation wireless systems can be viewed as a problem of structural robustness, namely the ability of the network to preserve its communication function under perturbations such as CSI uncertainty, residual successive interference cancellation (SIC) interference, spatial correlation, and infrastructure failures \cite{Cinkler2020,Abhishek2020}. Traditional resilience mechanisms have often emphasized redundancy, whereas modern wireless architectures increasingly rely on adaptive reconfiguration, virtualization, and resource-aware coordination to maintain service under changing operating conditions \cite{Cinkler2020,Abhishek2020}. From a system theoretic perspective, such resilience emerges from the availability of multiple functionally viable resource-allocation states rather than from exact duplication alone \cite{Neal2017,Merim2020}. This viewpoint is closely related to degeneracy, in which structurally distinct components or strategies realize similar functional outcomes under varying conditions \cite{Giulio1999}. Unlike redundancy, which replicates the same function through similar structures, degeneracy allows heterogeneous mechanisms to preserve functionality while retaining distinct operational characteristics, thereby improving adaptability and robustness \cite{Garima2023,Kibilda2025}. In recent years, semi-supervised learning approaches have gained attention for their ability to improve communication system performance, with notable work such as \cite{Guo2024SemiAMR} focusing on automatic modulation recognition with consistency regularization. The integration of energy harvesting techniques, such as time switching and power splitting, has been explored to enhance the reliability of overlay radio networks, as shown in \cite{Tashman2024Maximizing}.

Viewed from this perspective, RSMA is not only an interference-management technique but also a mechanism enlarging wireless network's functional state space. By enabling flexible common–private power allocation and decoding strategies, RSMA supports multiple feasible transmission configurations that satisfy the same QoS requirements under varying channel and interference conditions \cite{Zhou2021}. This structural flexibility enables more graceful performance degradation compared to rigid OMA or fixed non-orthogonal schemes, and highlights the benefit of RSMA under imperfect CSI extending beyond rate gains to enhanced adaptability and network resilience.

Interestingly, although degeneracy has not been explicitly formalized in the RSMA literature, prior results already suggest that multiple rate-splitting configurations and power-allocation structures can achieve comparable spectral-efficiency and energy-efficiency outcomes \cite{Zhou2021}. This multiplicity indicates that RSMA naturally supports functionally equivalent transmission states under different channel and interference realizations. Such structural flexibility can be exploited to design resilience-aware wireless systems that adapt access modes and power allocation in response to CSI imperfections, residual SIC interference, spatial correlation, and infrastructure failures. In this sense, OMA fallback is not merely a backup rule, but a complementary mechanism within a broader degeneracy-aware design framework that preserves the communication objective through a structurally different access strategy.
Motivated by this perspective, this work develops a degeneracy-aware, resilience-oriented transmission framework for multi-user multiple-input multiple-output (MIMO) networks operating under practical impairments. By jointly considering imperfect CSI, residual SIC interference, and base-station failures, the proposed framework integrates adaptive RSMA--OMA switching with robust power control to sustain performance across varying SNR regimes and network conditions. In doing so, it connects interference management with the broader question of how multiple functionally equivalent transmission states enable graceful service degradation rather than abrupt functional collapse in next-generation wireless systems.

\subsection*{Contributions}
This paper makes the following key contributions:
\begin{itemize}
    \item We propose a novel degeneracy-aware power control framework for RSMA in MIMO systems, addressing the challenges posed by imperfect CSI and residual SIC interference, ensuring system feasibility and adaptability under these impairments.
    \item We analyze the impact of CSI uncertainty and residual SIC interference on RSMA power efficiency and propose an adaptive fallback mechanism that switches to OMA whenever RSMA becomes infeasible or power-inefficient.
    \item We provide a comprehensive simulation-based performance evaluation, showing that the proposed framework significantly enhances system resilience, reduces outage probability, and improves overall power efficiency under real-world conditions.
    \item We validate the robustness of the proposed approach across various network scenarios, showcasing its ability to maintain high system performance despite the presence of CSI errors and residual SIC interference.
\end{itemize}
\begin{table}[t]
\caption{Key Symbols and Notations}
\label{tab:symbols}
\centering
\footnotesize
\setlength{\tabcolsep}{4pt}
\renewcommand{\arraystretch}{1.05}
\begin{tabularx}{\columnwidth}{>{\raggedright\arraybackslash}p{2.0cm} >{\raggedright\arraybackslash}X}
\hline
\textbf{Symbol} & \textbf{Description} \\
\hline
$M_t$ & Number of transmit antennas at the base station (BS) \\
$K$ & Number of users in the system \\
$\mathbf{H}_k(t)$ & Downlink channel matrix between the BS and user $k$ at time $t$ \\
$\hat{\mathbf{H}}_k(t)$ & Estimated channel matrix of user $k$ at the BS \\
$\epsilon_k^2$ & Normalized CSI error variance of user $k$ \\
$\mathbf{W}_c(t)$ & Precoder for the common RSMA stream \\
$\mathbf{W}_k(t)$ & Precoder for the private stream of user $k$ \\
$P_t$ & Total transmit power budget \\
$P_c(t)$ & Power allocated to the common stream \\
$P_k(t)$ & Power allocated to the private stream of user $k$ \\
$\sigma^2$ & Noise variance \\
$\gamma_{c,k}(t)$ & SINR for decoding the common stream at user $k$ \\
$\gamma_{p,k}(t)$ & SINR for decoding the private stream at user $k$ \\
$\xi_k$ & Residual SIC interference factor \\
$r_k^{\mathrm{RSMA}}(t)$ & Achievable rate of user $k$ under RSMA \\
$r_k^{\mathrm{OMA}}(t)$ & Achievable rate of user $k$ under OMA \\
$r_{t,k}$ & Target rate requirement of user $k$ \\
$a_{\mathrm{RSMA}}(t)$ & RSMA mode indicator (1: RSMA, 0: OMA) \\
$P_{\mathrm{out}}^{\mathrm{RSMA}}(t)$ & Outage probability under RSMA \\
$P_{\mathrm{th}}$ & Maximum tolerable outage probability \\
$\mathcal{R}_{\mathrm{BS}}(t)$ & Base station resilience indicator \\
\hline
\end{tabularx}
\end{table}

\begin{figure*}[t]
    \centering
\includegraphics[width=0.7\linewidth]{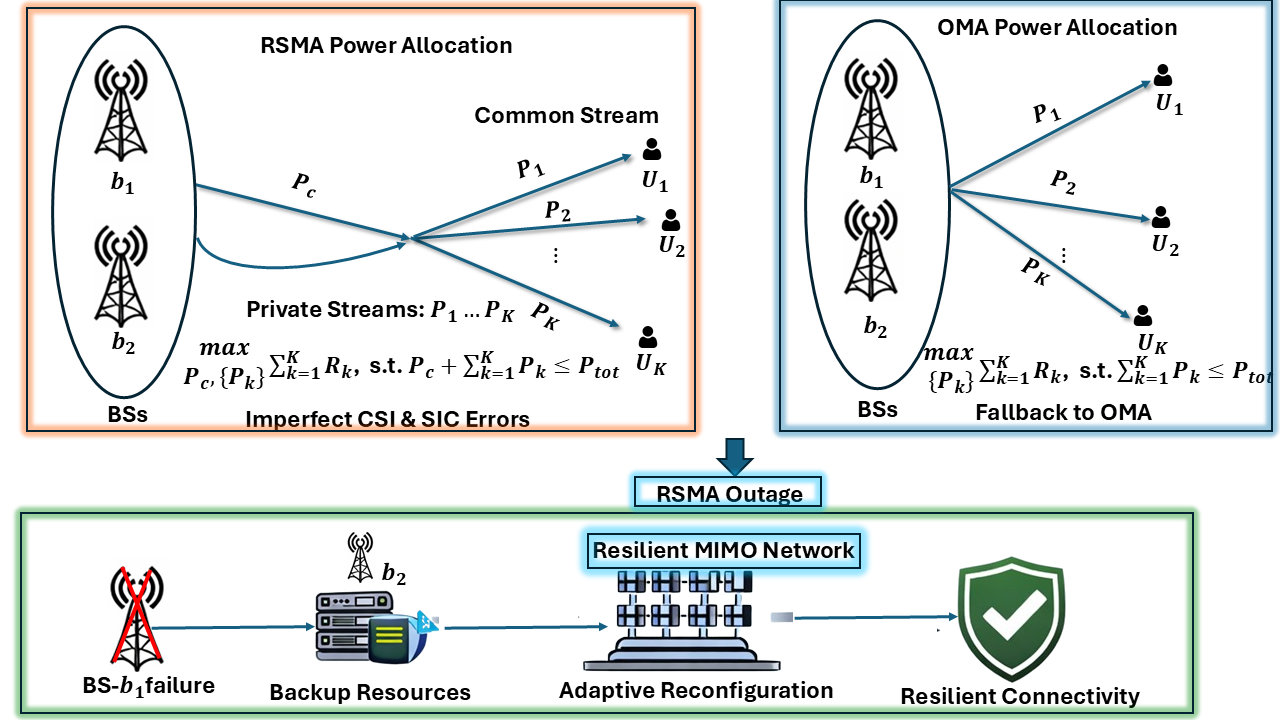}
    \caption{Resilient multi-user MIMO system model comparing RSMA and OMA power allocation. The framework captures (i) joint transmission from BS-$b_1$ and BS-$b_2$ under normal conditions and (ii) network resilience under BS-$b_1$ failure, where BS-$b_2$ dynamically reallocates resources to maintain connectivity, highlighting robustness against imperfect CSI and SIC interference.}
    \label{system model}
    \vspace{-4mm}
    \end{figure*}

\section{System Model}
We consider a downlink single-cell multiuser MIMO system in which a base station (BS) equipped with $M_t$ transmit antennas serves $K$ users, each with $M_r$ receive antennas as shown in Fig. \ref{system model}. The BS employs RSMA as the primary transmission strategy. For user $k$, the message is divided into a common part and a private part. The common parts of all users are jointly encoded into one common stream, which is decoded by all users, while the private part of each user is encoded into an individual private stream intended only for that user. A summary of the main symbols used is provided in Table~\ref{tab:symbols}. \vspace{-4mm}

\subsection{Degeneracy-Aware RSMA Operation and Fallback Principle}

The BS allocates power over one common stream and $K$ private streams subject to the total transmit-power constraint and the users' target-rate requirements. Under strong spatial correlation, imperfect CSI, and residual SIC interference, RSMA may remain feasible only for a narrow set of power allocations. To formalize this structural flexibility, we define degeneracy as the multiplicity of distinct power-allocation states that achieve the same functional objective, namely satisfaction of the target quality-of-service (QoS) vector.

Let $\mathbf{p}(t)=[P_c(t),P_1(t),\ldots,P_K(t)]^T\in\mathbb{R}_+^{K+1}$ denote an RSMA power-allocation state at time $t$. The corresponding achieved rate vector is given by $\mathbf{r}^{\mathrm{RSMA}}(\mathbf{p}(t),t)=[r^{\mathrm{RSMA}}_1(\mathbf{p}(t),t),\ldots,r^{\mathrm{RSMA}}_K(\mathbf{p}(t),t)]^T$.

Let $\mathbf{r}_t=[r_{t,1},\ldots,r_{t,K}]^T$ denote the target-rate vector. 
Let $P_{\mathrm{RSMA}}(t)$ denote the minimum total transmit power required by RSMA to satisfy the QoS constraints. Specifically,
$P_{\mathrm{RSMA}}(t)=\min_{\mathbf{p}(t)\in\mathbb{R}_+^{K+1}}\mathbf{1}^T\mathbf{p}(t)$, where $\mathbf{1}$ denotes a $(K+1)\times 1$ column vector of ones.
The constraints are $r_k^{\mathrm{RSMA}}(\mathbf{p}(t),t)\ge r_{t,k}$ for all $k$.

To quantify the number of functionally equivalent operating points, we introduce a power quantization step $\Delta>0$. The admissible power grid is defined as $\mathcal{Q}_{\Delta}(t)=\{\mathbf{p}(t)\in\Delta\mathbb{Z}_+^{K+1}:\mathbf{1}^T\mathbf{p}(t)\le P_t\}$, where $P_t$ denotes the available transmit-power budget at time $t$. This quantization converts the continuous feasible region into a countable set of operating states.

For a design margin $\varepsilon_P\ge0$, the set of QoS-equivalent RSMA states is defined as $\mathcal{F}_{\mathrm{eq}}^{\mathrm{RSMA}}(t)=\{\mathbf{p}(t)\in\mathcal{Q}_{\Delta}(t):r_k^{\mathrm{RSMA}}(\mathbf{p}(t),t)\ge r_{t,k},\ \forall k,\ \mathbf{1}^T\mathbf{p}(t)\le P_{\mathrm{RSMA}}(t)+\varepsilon_P\}$.

The set $\mathcal{F}_{\mathrm{eq}}^{\mathrm{RSMA}}(t)$ includes all structurally distinct RSMA power allocations that satisfy the same QoS requirement while remaining within an acceptable power margin of the minimum-power solution. Hence, these states are functionally equivalent in terms of QoS support. The number of functionally equivalent RSMA states is then given by $N_{\mathrm{eq}}^{\mathrm{RSMA}}(t)
=
\left|
\mathcal{F}_{\mathrm{eq}}^{\mathrm{RSMA}}(t)
\right|$, and the corresponding Degeneracy Grade is defined as
\begin{equation}
D_{\mathrm{deg}}^{\mathrm{RSMA}}(t)
=
\frac{
\log\!\left(1+N_{\mathrm{eq}}^{\mathrm{RSMA}}(t)\right)
}{
\log\!\left(1+\left|\mathcal{Q}_{\Delta}(t)\right|\right)
}
\in [0,1].
\end{equation}
This normalized metric measures the density of QoS-equivalent RSMA operating points within the admissible power space. A larger value of \(D_{\mathrm{deg}}^{\mathrm{RSMA}}(t)\) indicates that the target QoS can be sustained by many distinct RSMA power-allocation states, which implies greater structural flexibility and stronger tolerance to CSI perturbations, residual SIC interference, and local operating-point variations. In contrast, \(D_{\mathrm{deg}}^{\mathrm{RSMA}}(t)=0\) implies that no quantized RSMA state satisfies the target QoS within the admissible power margin. This formalization also clarifies the role of fallback. When the RSMA feasible set is empty, or when its degeneracy grade becomes very small under severe uncertainty, the system loses structurally distinct RSMA alternatives for maintaining the required QoS. In that regime, OMA acts as a functionally equivalent fallback mechanism at the access-strategy level. Therefore, degeneracy is not used only as a conceptual justification for switching. It is quantified through the cardinality of the QoS-equivalent feasible set, which directly measures how many alternative power-allocation states are available to maintain the same communication objective under a given channel realization. \vspace{-3mm} 

\subsection{Spatially Correlated MIMO Channel With Imperfect CSI}

At time instant $t$, the downlink channel matrix from the BS to user $k$ is modeled as $\mathbf{H}_k(t)
=\sqrt{\beta_k(t)}\, \mathbf{R}_{r,k}^{1/2}
\mathbf{G}_k(t)
\mathbf{R}_{t,k}^{1/2}$, where $\beta_k(t)=d_k(t)^{-\eta}$ is the large-scale path-loss coefficient, $d_k(t)$ denotes the distance between the BS and user $k$, and $\eta$ is the path-loss exponent. The matrix $\mathbf{G}_k(t)\in\mathbb{C}^{M_r\times M_t}$ models small-scale fading and has independent and identically distributed entries drawn from $\mathcal{CN}(0,1)$. The matrices $\mathbf{R}_{r,k}$ and $\mathbf{R}_{t,k}$ denote the receive-side and transmit-side spatial correlation matrices, respectively. The BS is assumed to have imperfect CSI due to estimation errors and feedback limitations. Accordingly, the instantaneous channel is decomposed as
$\mathbf{H}_k(t)=\hat{\mathbf{H}}_k(t)+\tilde{\mathbf{H}}_k(t)$, where $\hat{\mathbf{H}}_k(t)$ is the channel estimate available at the BS and $\tilde{\mathbf{H}}_k(t)$ is the corresponding estimation error. We assume that the estimated channel and the estimation error follow the same spatial correlation structure. Their distributions are given by
$\hat{\mathbf{H}}_k(t)\sim\mathcal{CN}(\mathbf{0},(1-\epsilon_k^2)\beta_k(t)\mathbf{R}_{r,k}\otimes\mathbf{R}_{t,k}),\ 
\tilde{\mathbf{H}}_k(t)\sim\mathcal{CN}(\mathbf{0},\epsilon_k^2\beta_k(t)\mathbf{R}_{r,k}\otimes\mathbf{R}_{t,k}).$
where $\epsilon_k^2\in[0,1]$ is the normalized CSI error variance. The case $\epsilon_k^2=0$ corresponds to perfect CSI, whereas larger values of $\epsilon_k^2$ indicate more severe channel uncertainty. The Kronecker product $\otimes$ is used to represent separable spatial correlation across the transmit and receive antenna arrays. This channel model captures three key impairments relevant to the proposed framework: path loss through $\beta_k(t)$, spatial correlation through $\mathbf{R}_{t,k}$ and $\mathbf{R}_{r,k}$, and CSI uncertainty through $\epsilon_k^2$. These factors directly affect the feasibility and efficiency of RSMA power allocation and therefore play a central role in the subsequent access-mode selection framework. \vspace{-4mm}

\subsection{RSMA Transmission Model}

At time instant $t$, the BS transmits one common stream and $K$ private streams. Under RSMA, the transmitted signal is given by \cite{Zhou2021}.
The signal is expressed as $\mathbf{x}(t)=\sqrt{P_c(t)}\,\mathbf{W}_c(t)\mathbf{s}_c(t)+\sum_{k=1}^{K}\sqrt{P_k(t)}\,\mathbf{W}_k(t)\mathbf{s}_k(t)$. Here, $\mathbf{s}_c(t)$ and $\mathbf{s}_k(t)$ denote the common and private data streams, respectively. The precoding matrices $\mathbf{W}_c(t)\in\mathbb{C}^{M_t\times d_c}$ and $\mathbf{W}_k(t)\in\mathbb{C}^{M_t\times d_k}$ correspond to the common and private streams. The parameters $d_c$ and $d_k$ denote the corresponding stream dimensions, and the allocated powers are $P_c(t)$ and $P_k(t)$. The BS is subject to the total transmit-power constraint $P_c(t)+\sum_{k=1}^{K}P_k(t)\le P_{\max}$, where $P_{\max}$ denotes the maximum available transmit power per time slot. The received signal at user $k$ is
$\mathbf{y}_k(t)=\mathbf{H}_k(t)\mathbf{x}(t)+\mathbf{n}_k(t)$,
where $\mathbf{H}_k(t)\in\mathbb{C}^{M_r\times M_t}$ denotes the downlink channel matrix, and $\mathbf{n}_k(t)\sim\mathcal{CN}(\mathbf{0},\sigma^2\mathbf{I}_{M_r})$ is the additive white Gaussian noise vector. For notational convenience, the transmitted streams are assumed to be independent and normalized. In particular, $\mathbb{E}[\mathbf{s}_c(t)\mathbf{s}_c^{H}(t)]=\mathbf{I}_{d_c}$ and $\mathbb{E}[\mathbf{s}_k(t)\mathbf{s}_k^{H}(t)]=\mathbf{I}_{d_k}$ for all $k$. Moreover, $\mathbb{E}[\mathbf{s}_c(t)\mathbf{s}_k^{H}(t)]=\mathbf{0}$ and $\mathbb{E}[\mathbf{s}_i(t)\mathbf{s}_j^{H}(t)]=\mathbf{0}$ for all $i\neq j$. Under RSMA, each user first decodes the common stream while treating all private streams as interference. After decoding the common stream, the receiver applies SIC before decoding its intended private stream. \vspace{-4mm}

\subsection{Achievable Rates Under Imperfect SIC}

Under RSMA, each user decodes the common stream first and then decodes its intended private stream. When decoding the common stream, all private streams are treated as interference. Accordingly, the instantaneous SINR for decoding the common stream at user $k$ is given by
\begin{equation}
\label{eq:8_time}
\gamma_{c,k}(t)
=
\frac{P_c(t)\left\|\mathbf{H}_k(t)\mathbf{W}_c(t)\right\|_F^2}
{\sum_{j=1}^{K} P_j(t)\left\|\mathbf{H}_k(t)\mathbf{W}_j(t)\right\|_F^2
+ \sigma^2}.
\end{equation}

Since the common stream must be decoded successfully by all users, its achievable rate is determined by the worst-user SINR, namely
$r_c(t)
=
\min_{k\in\{1,\dots,K\}}
\log_2\!\left(1+\gamma_{c,k}(t)\right)$. After decoding the common stream, user $k$ applies SIC before decoding its private stream. To account for imperfect SIC, we let $\xi_k(t)\in[0,1]$ denote the residual SIC interference factor at user $k$, where $\xi_k(t)=0$ corresponds to perfect SIC and larger values indicate stronger residual interference from the common stream. The resulting SINR for decoding the private stream of user $k$ is
\begin{align}
\label{eq:10_time}
\gamma_{p,k}(t) &= \frac{A}{\sum_{j\neq k} B+C+\sigma^2} 
\end{align}
where
\begin{align*}
A &= P_k(t)\|\mathbf{H}_k(t)\mathbf{W}_k(t)\|_F^2,\\
B &= P_j(t)\|\mathbf{H}_k(t)\mathbf{W}_j(t)\|_F^2,\\
C &= \xi_k(t)P_c(t)\|\mathbf{H}_k(t)\mathbf{W}_c(t)\|_F^2.
\end{align*}

The corresponding achievable private rate of user $k$ is
$r_{p,k}(t) = \log_2\!\left(1+\gamma_{p,k}(t)\right)$. The total achievable RSMA rate of user $k$ is then given by
$
r_k^{\mathrm{RSMA}}(t)
=
c_k(t)+r_{p,k}(t)$,
where $c_k(t)\geq 0$ denotes the portion of the common rate assigned to user $k$. The common-rate allocation must satisfy $\sum_{k=1}^{K} c_k(t)\leq r_c(t)$. Therefore, the achievable rate of each user consists of two components: an allocated portion of the common stream and the rate of its private stream. These rate expressions form the basis for the feasibility and power-allocation framework developed in the next subsection.
\vspace{-9mm}
\subsection{Degeneracy-Aware RSMA--OMA Switching}
To maintain robust operation under time-varying channel conditions and receiver impairments, the BS adaptively selects between RSMA and OMA at each time instant. The switching decision is based on two criteria: 1) whether RSMA can satisfy the target-rate constraints within the available power budget, and 2) whether its instantaneous outage ratio remains below a prescribed threshold. Let $P_{\mathrm{RSMA}}^{\mathrm{req}}(t)$ denote the minimum total transmit power required by RSMA to satisfy the target-rate constraints at time $t$, and let $P_{\mathrm{OMA}}^{\mathrm{req}}(t)$ denote the corresponding minimum power required under OMA. The RSMA activation indicator is defined as $a_{\mathrm{RSMA}}(t) =
\mathbb{1}
\left\{
P_{\mathrm{RSMA}}^{\mathrm{req}}(t)\le P_t
\;\land\;
P_{\mathrm{out}}^{\mathrm{RSMA}}(t)\le P_{\mathrm{th}}
\right\}$, where $\mathbb{1}\{\cdot\}$ denotes the indicator function, which equals 1 if the condition inside $\{\cdot\}$ is satisfied and 0 otherwise, $P_t$ is the available transmit-power budget at time $t$, and $P_{\mathrm{th}}$ is the maximum tolerable outage ratio. This rule reflects the degeneracy-aware fallback principle adopted in this paper: RSMA is used when it remains feasible and reliable, while OMA serves as an alternative mode when RSMA becomes infeasible or outage-prone. In this sense, RSMA and OMA are treated as distinct transmission strategies capable of fulfilling the same communication objective under different operating conditions. When RSMA is not selected, the system switches to OMA. The achievable rate of user $k$ under OMA is modeled as
$r_k^{\mathrm{OMA}}(t)
=
\tau_k(t)\log_2\!\left(
1+
\frac{P_k^{\mathrm{OMA}}(t)\|\mathbf{H}_k(t)\|_F^2}{\sigma^2}
\right)$, subject to $\sum_{k=1}^{K}\tau_k(t)\le 1, \qquad \sum_{k=1}^{K}P_k^{\mathrm{OMA}}(t)\le P_t$, where $\tau_k(t)$ denotes the fraction of orthogonal resources assigned to user $k$. Accordingly, the effective achievable rate of user $k$ at time $t$ is
$r_k(t)
=
a_{\mathrm{RSMA}}(t)\,r_k^{\mathrm{RSMA}}(t)
+\bigl(1-a_{\mathrm{RSMA}}(t)\bigr)\,r_k^{\mathrm{OMA}}(t)$.\vspace{-3mm}

\section{Feasibility-Driven Power Allocation and Performance Metrics}

This section develops the proposed RSMA power-allocation framework, establishes the corresponding feasibility conditions, and defines the performance metrics used for evaluation. For a given channel realization at time instant $t$, the precoding matrices are assumed to be fixed, and the design focuses on power allocation, common-rate splitting, and access-mode selection.\vspace{-5mm}

\subsection{Minimum-Power RSMA Design}

Given the target-rate vector $\{r_{t,k}\}_{k=1}^{K}$, the RSMA design problem at time instant $t$ is formulated as a minimum-power feasibility problem. The objective is to determine the smallest transmit power required to satisfy all user rate constraints under the RSMA transmission model. The problem is written as
\begin{align}
\label{eq:fp1}
\min_{P_c(t),\,\{P_k(t)\},\,\{c_k(t)\}} \quad
& P_c(t)+\sum_{k=1}^{K}P_k(t) \\
\label{eq:fp2}
\text{s.t.}\quad
& c_k(t)+r_{p,k}(t)\ge r_{t,k}, \quad \forall k, \\
\label{eq:fp3}
& \sum_{k=1}^{K} c_k(t)\le r_c(t), \\
\label{eq:fp4}
& P_c(t)\ge 0,\; P_k(t)\ge 0,\quad \forall k, \\
\label{eq:fp5}
& c_k(t)\ge 0,\quad \forall k.
\end{align}
Let $P_{\mathrm{RSMA}}^{\mathrm{req}}(t)$ denote the optimal value of eqs. \eqref{eq:fp1}-\eqref{eq:fp5}. This quantity represents the minimum total power required by RSMA to support the target-rate vector at time $t$. The coupling between SINR values and common/private rate constraints renders the RSMA feasibility problem nonconvex (Appendix~A), which motivates the adoption of the proposed sequential heuristic approach.\vspace{-4mm}
\subsection{Structure of the Optimal Solution}
Minimizing the total transmit power subject to target-rate constraints, as in eq. \eqref{eq:fp1}, ensures that any excess rate allocation is suboptimal. Therefore, at optimality, each user's total rate constraint is active:
$r_k^{\mathrm{RSMA}}(t) = r_{t,k}, \ \forall k$. This implies that the common-rate allocation and private-rate allocation must jointly satisfy each user's target rate with equality. Accordingly, the optimal common-rate allocation is
$c_k^\star(t) = \max\bigl(0,\, r_{t,k} - r_{p,k}(t)\bigr), \ \forall k$. For a given common-rate allocation, the required private-stream SINR of user $k$ is
$\Gamma_{p,k}(t) = 2^{r_{t,k} - c_k^\star(t)} - 1$. Since the objective is to minimize power, the private-stream SINR constraint must also hold with equality at optimum. Substituting eq. \eqref{eq:10_time} into $\gamma_{p,k}(t)=\Gamma_{p,k}(t)$ yields
\begin{align}
\label{eq:25}
P_k^\star(t)
&=
\frac{\Gamma_{p,k}(t)}
{\|\mathbf{H}_k(t)\mathbf{W}_k(t)\|_F^2}
\Bigg(
\sum_{j\neq k} P_j^\star(t)\|\mathbf{H}_k(t)\mathbf{W}_j(t)\|_F^2 \nonumber\\
&+
\xi_k(t)P_c^\star(t)\|\mathbf{H}_k(t)\mathbf{W}_c(t)\|_F^2
+
\sigma^2
\Bigg),
\end{align}
for all $k\in\{1,\dots,K\}$. Eq. \eqref{eq:25} gives the power required by the private stream of user $k$ in the presence of multiuser interference, residual SIC interference, and noise.\vspace{-4mm}
\subsection{Common-Stream Power Requirement}
The common stream must support the total common rate assigned across all users. Therefore, for every user $k$, the common-stream SINR must satisfy
\begin{equation}
\label{eq:26}
\gamma_{c,k}(t)
\ge
2^{\sum_{j=1}^{K}c_j^\star(t)}-1.
\end{equation}
Again, since the objective is minimum-power transmission, the above constraint is active at the optimum. Substituting eq. \eqref{eq:8_time} into eq. \eqref{eq:26} gives the required common-stream power
\begin{align}
\label{eq:27}
P_c^\star(t)
&=
\max_{k\in\{1,\dots,K\}}
\left(2^{\sum_{j=1}^{K}c_j^\star(t)}-1\right)\nonumber\\
&\frac{\left(
\sum_{j=1}^{K}P_j^\star(t)\|\mathbf{H}_k(t)\mathbf{W}_j(t)\|_F^2+\sigma^2
\right)
}{
\|\mathbf{H}_k(t)\mathbf{W}_c(t)\|_F^2
}.
\end{align}

Eq. \eqref{eq:27} shows that the common-stream power is determined by the most demanding user, i.e., the user requiring the largest common-stream power to decode the aggregate common message.\vspace{-4mm}
\subsection{RSMA Feasibility Condition}
The RSMA design is feasible at time instant $t$ only if the required total power does not exceed the available power budget. Hence, the feasibility condition is
\begin{equation}
\label{eq:28}
P_c^\star(t)+\sum_{k=1}^{K}P_k^\star(t)\le P_t.
\end{equation}
If eq. \eqref{eq:28} holds, then RSMA can satisfy all target-rate constraints at time $t$. Otherwise, no RSMA power allocation exists that simultaneously meets the target rates and the transmit-power budget under the given channel conditions, CSI uncertainty, spatial correlation, and residual SIC interference level. Accordingly, the minimum required power under RSMA is defined as $P_{\mathrm{RSMA}}^{\mathrm{req}}(t) = P_c^\star(t)+\sum_{k=1}^{K}P_k^\star(t)$. The RSMA power allocation constitutes a coupled fixed-point system (Appendix~D), thereby justifying the use of the iterative feasibility-based solver employed in Algorithm~\ref{alg:degeneracy_switching}.\vspace{-4mm}
\subsection{Outage Probability}
At time instant $t$, user $k$ is in outage under RSMA if its achievable rate falls below its target rate. The corresponding outage indicator is
$\mathcal{I}_{k}^{\mathrm{RSMA}}(t)
=\mathbb{1} \left\{
r_k^{\mathrm{RSMA}}(t)<r_{t,k}
\right\}$.
The instantaneous RSMA outage ratio is defined as
$P_{\mathrm{out}}^{\mathrm{RSMA}}(t)=\frac{1}{K}\sum_{k=1}^{K}\mathcal{I}_{k}^{\mathrm{RSMA}}(t)$.
The average outage probability over a horizon of $T$ time slots is

\begin{equation}
\label{eq:13b}
\bar{P}_{\mathrm{out}}^{\mathrm{RSMA}}
=
\frac{1}{KT}
\sum_{t=1}^{T}\sum_{k=1}^{K}
\mathcal{I}_{k}^{\mathrm{RSMA}}(t).
\end{equation}

The instantaneous outage ratio is used for access-mode selection, while eq. \eqref{eq:13b} is used as a long-term performance measure. \vspace{-4mm}

\subsection{Average Sum Rate and Throughput Before BS Failure}
Before any BS failure event, the average sum rate under adaptive access-mode selection is defined as
\begin{align}
\label{eq:avg_sum_rate}
\bar{R}^{\mathrm{NR}}_{\mathrm{sum}}
&=
\mathbb{E}
\bigg[
\sum_{k=1}^{K}
\big(
a_{\mathrm{RSMA}}(t)\,r_k^{\mathrm{RSMA}}(t)\nonumber\\
&+
\big(1-a_{\mathrm{RSMA}}(t)\big)\,r_k^{\mathrm{OMA}}(t)
\big)
\bigg],
\end{align}

where the expectation is taken over the spatially correlated fading channels, CSI estimation errors, and residual SIC interference effects. To account for both the achieved rate and reliability, the instantaneous system throughput before BS failure is defined as
\begin{align}
T_{\mathrm{sys}}^{\mathrm{NR}}(t)
&=
\left(1-P_{\mathrm{out}}^{\mathrm{RSMA}}(t)\right)
\sum_{k=1}^{K}
\Big(
a_{\mathrm{RSMA}}(t)\,r_k^{\mathrm{RSMA}}(t)
\nonumber\\
&\quad+
\bigl(1-a_{\mathrm{RSMA}}(t)\bigr)\,r_k^{\mathrm{OMA}}(t)
\Big).
\end{align}
The corresponding time-averaged throughput over $T$ time slots:

\begin{equation}
\label{eq:throughput_avg_time}
\bar{T}_{\mathrm{sys}}^{\mathrm{NR}}
=
\frac{1}{T}
\sum_{t=1}^{T}
T_{\mathrm{sys}}^{\mathrm{NR}}(t).
\end{equation}

These metrics quantify the long-term performance of the proposed adaptive framework before any infrastructure failure occurs.
\begin{algorithm}[t]
\caption{Feasibility-Driven Degeneracy-Aware RSMA--OMA Switching}
\label{alg:degeneracy_switching}
\footnotesize
\begin{algorithmic}[1]
\STATE \textbf{Input:} $\{\hat{\mathbf{H}}_k(t)\}$, $\mathbf{W}_c(t)$, $\{\mathbf{W}_k(t)\}$, $\{r_{t,k}\}$, $P_t$, $\sigma^2$, $\{\xi_k(t)\}$, $P_{\mathrm{th}}$

\STATE Compute effective gains:
\[
g_{c,k}(t)=\|\mathbf{H}_k(t)\mathbf{W}_c(t)\|_F^2,\quad
g_{k,j}(t)=\|\mathbf{H}_k(t)\mathbf{W}_j(t)\|_F^2
\]

\STATE \textbf{RSMA feasibility:}
\FOR{$k=1$ to $K$}
\STATE Compute $r_{p,k}(t)$, $c_k^\star(t)=\max(0,r_{t,k}-r_{p,k}(t))$, and $\Gamma_{p,k}(t)=2^{r_{t,k}-c_k^\star(t)}-1$
\ENDFOR

\STATE Solve $\{P_k^\star(t)\}$ from:
\[
P_k^\star(t)=
\frac{\Gamma_{p,k}(t)}{g_{k,k}(t)}
\left(
\sum_{j\neq k}P_j^\star(t)g_{k,j}(t)
+\xi_k(t)P_c^\star(t)g_{c,k}(t)
+\sigma^2
\right)
\]

\STATE Compute:
\[
P_c^\star(t)=
\max_k
\frac{
\left(2^{\sum_j c_j^\star(t)}-1\right)
\left(\sum_j P_j^\star(t)g_{k,j}(t)+\sigma^2\right)
}{
g_{c,k}(t)
}
\]

\STATE Compute $P_{\mathrm{RSMA}}^{\mathrm{req}}(t)=P_c^\star(t)+\sum_k P_k^\star(t)$ and
\[
P_{\mathrm{out}}^{\mathrm{RSMA}}(t)=
\frac{1}{K}\sum_k
\mathbb{1}\{r_k^{\mathrm{RSMA}}(t)<r_{t,k}\}
\]

\STATE \textbf{OMA feasibility:} Compute $P_{\mathrm{OMA}}^{\mathrm{req}}(t)$

\STATE \textbf{Mode selection:}
\IF{$P_{\mathrm{RSMA}}^{\mathrm{req}}(t)\le P_t$ and $P_{\mathrm{out}}^{\mathrm{RSMA}}(t)\le P_{\mathrm{th}}$}
\STATE $a_{\mathrm{RSMA}}(t)=1$; allocate $\{P_c^\star(t),P_k^\star(t),c_k^\star(t)\}$
\ELSE
\STATE $a_{\mathrm{RSMA}}(t)=0$; use OMA if $P_{\mathrm{OMA}}^{\mathrm{req}}(t)\le P_t$, else declare outage
\ENDIF

\STATE \textbf{Output:} $a_{\mathrm{RSMA}}(t)$ and corresponding allocation
\end{algorithmic}
\end{algorithm}
Because the RSMA feasibility problem in eqs. \eqref{eq:fp1}-\eqref{eq:fp5} is nonconvex due to the coupled SINR, common-rate, and power variables, Algorithm~\ref{alg:degeneracy_switching} should be interpreted as a structured sequential feasibility solver rather than a globally optimal nonconvex optimizer. In particular, the algorithm exploits the fixed-point structure induced by eq. \eqref{eq:25} and eq. \eqref{eq:27} for a given common-rate allocation and then combines this RSMA feasibility test with the OMA fallback rule. Hence, the resulting solution is a high-quality feasible operating point for online mode selection, but global optimality for the original joint problem is not claimed. In contrast, the OMA fallback problem possesses a convex structure (Appendix~B), which enables efficient computation of feasible power allocations under orthogonal resource constraints.\vspace{-4mm}
\subsection{Optimality Gap and Computational Complexity of Algorithm~\ref{alg:degeneracy_switching}}
The non-convexity established for eq. \eqref{eq:fp1}-\eqref{eq:fp5} implies that Algorithm~\ref{alg:degeneracy_switching} does not, in general, guarantee the global minimum-power solution of the original RSMA problem. More precisely, the bilinear dependence of the private-stream SINR targets on the common-rate variables, together with the interference coupling among the power variables, prevents a direct claim of global optimality. Therefore, Algorithm~\ref{alg:degeneracy_switching} is best interpreted as a sequential heuristic that exploits the analytical structure of eq. \eqref{eq:25} and eq. \eqref{eq:27} to construct a feasible RSMA operating point whenever one is found, and then performs access-mode switching according to the outage and power criteria. Let \(P_{\rm RSMA}^{\rm Alg}(t)\) denote the RSMA power returned by Algorithm~\ref{alg:degeneracy_switching} and let \(P_{\rm RSMA}^{\rm Glob}(t)\) denote the globally optimal value of eq. \eqref{eq:fp1}-\eqref{eq:fp5}, obtained only for benchmarking through an offline global solver such as branch-and-bound. The relative optimality gap is defined as
$
\Delta_{\rm opt}(t) = \frac{P_{\rm RSMA}^{\rm Alg}(t) - P_{\rm RSMA}^{\rm Glob}(t)}{\max\{P_{\rm RSMA}^{\rm Glob}(t), \epsilon_0\}}$,
where \(\epsilon_0 > 0\) is a small numerical constant introduced to avoid division by zero. By construction, \(\Delta_{\rm opt}(t) \ge 0\), and \(\Delta_{\rm opt}(t) = 0\) only when the algorithm matches the global optimum. In the simulation study, this metric can be evaluated for small user sets in order to quantify the price paid for engineering simplicity. The computational advantage of Algorithm~\ref{alg:degeneracy_switching} follows from its structured decomposition. For each time instant, the dominant operations are: 1) evaluation of the effective gains \(\{g_{c,k}(t), g_{k,j}(t)\}\), which requires \(O(K^2)\) scalar gain computations once the precoders are fixed; 2) iterative solution of the coupled private and common power updates in eq. \eqref{eq:25} and eq. \eqref{eq:27}; and 3) evaluation of the OMA fallback feasibility problem. If \(I_{\rm fp}\) denotes the number of fixed-point iterations required for convergence of the RSMA updates, then the overall complexity of the proposed switching rule is $\mathcal{C}_{\rm Alg} = O(I_{\rm fp} K^2) + \mathcal{C}_{\rm OMA}$, 
where \(\mathcal{C}_{\rm OMA}\) is polynomial and remains modest because the OMA fallback problem is convex under the formulation in Appendix C. Hence, the per-slot complexity of the proposed degeneracy-aware switching remains polynomial in the number of users.
By contrast, a global branch-and-bound solver must explore the nonconvex search space associated with the joint variable tuple
\(
(P_c(t),\{P_k(t)\},\{c_k(t)\}),
\)
whose dimension is \(2K+1\). As a result, its worst-case complexity grows exponentially with the number of optimization variables. This can be expressed generically as
$\mathcal{C}_{\rm B\&B} =
O\!\left(B^{\,2K+1}\right)$, where \(B\) denotes the effective branching factor or resolution level of the global search. Therefore, although branch-and-bound is useful as an offline reference for small-scale instances, it is not suitable for fast online access-mode selection in time-varying channels. This comparison highlights the intended engineering role of Algorithm 1. The proposed method sacrifices global optimality guarantees in exchange for tractable online implementation, explicit feasibility interpretation, and seamless integration with the OMA fallback rule. Consequently, the relevant performance criterion is not exact global optimality alone, but the trade-off between solution quality and computational efficiency, which can be quantified through \(\Delta_{\rm opt}(t)\) together with the measured runtime reduction relative to the global benchmark.

\section{Resilience Under Base-Station Failure}

To evaluate service continuity under infrastructure disruption, we consider a two-BS deployment with
\(
\mathcal{B}=\{b_1,b_2\}
\).
Before failure, both BSs are operational and jointly support the user population. At the failure time \(t=t_f\), BS-\(b_1\) becomes unavailable. For \(t\ge t_f\), BS-\(b_2\) remains the only active transmitter and must serve all users subject to its own power budget and the prevailing channel conditions. The purpose of this model is to assess whether the proposed adaptive RSMA--OMA framework can preserve service continuity after the loss of one BS. In particular, after the failure of BS-\(b_1\), the surviving BS applies the same feasibility-driven access-mode selection strategy developed in the previous section. Therefore, at each time instant, BS-\(b_2\) selects RSMA when it can satisfy the target-rate requirements within the available power budget and outage threshold; otherwise, it switches to OMA whenever an OMA-feasible solution exists. \vspace{-3mm}

\subsection{Post-Failure Resilience Condition}

Let \(r_k^{(2)}(t)\) denote the effective achievable rate of user \(k\) when all users are served by BS-\(b_2\) after the failure event. The network is said to be resilient at time instant \(t\ge t_f\) if the surviving BS can support all target-rate requirements under the selected access mode. The corresponding resilience indicator is defined as
\begin{equation}
\label{eq:RBS_2BS}
\mathcal{R}_{\mathrm{BS}}(t)
=
\mathbb{1}
\left\{
r_k^{(2)}(t)\ge r_{t,k},\ \forall k
\right\},
\qquad t\ge t_f,
\end{equation}

where
\begin{equation}
\label{eq:rk_mode_b2}
\begin{aligned}
r_k^{(2)}(t)
&= a_{\mathrm{RSMA}}^{(2)}(t)\, r_k^{(2),\mathrm{RSMA}}(t) \\
&\quad + \left(1-a_{\mathrm{RSMA}}^{(2)}(t)\right)\, r_k^{(2),\mathrm{OMA}}(t).
\end{aligned}
\end{equation}

Here, \(a_{\mathrm{RSMA}}^{(2)}(t)\in\{0,1\}\) is the access-mode indicator of the surviving BS. A value of \(\mathcal{R}_{\mathrm{BS}}(t)=1\) indicates that BS-\(b_2\) can maintain service continuity for all users after the failure of BS-\(b_1\). In contrast, \(\mathcal{R}_{\mathrm{BS}}(t)=0\) indicates that no feasible post-failure transmission configuration can satisfy the required quality-of-service constraints. To make the feasibility condition explicit, the post-failure RSMA activation indicator at BS-\(b_2\) is defined, 
\begin{align}
\label{eq:arsma_b2}
a_{\mathrm{RSMA}}^{(2)}(t)
&=
\mathbb{1}
\left\{
P_{\mathrm{RSMA}}^{(2),\mathrm{req}}(t)\le P_t^{(2)}
\right.
\nonumber\\
&\qquad \left.
\text{and}\;
P_{\mathrm{out}}^{(2),\mathrm{RSMA}}(t)\le P_{\mathrm{th}}
\right\},
\end{align}
where \(P_{\mathrm{RSMA}}^{(2),req}(t)\) is the minimum power required by BS-\(b_2\) to satisfy the target-rate constraints under RSMA, \(P_t^{(2)}\) is the available transmit-power budget of BS-\(b_2\), and \(P_{\mathrm{out}}^{(2),\mathrm{RSMA}}(t)\) is the corresponding instantaneous outage ratio.

This formulation emphasizes that post-failure resilience is not tied to RSMA alone. Rather, resilience is achieved whenever the surviving BS can continue serving all users under either RSMA or OMA, depending on which mode remains feasible and reliable after the failure event. \vspace{-4mm}
\subsection{Average Sum Rate With Resilience}
To capture the impact of the failure event on system performance, we define a time-dependent average sum rate that distinguishes between pre-failure and post-failure operation. Specifically,
\begin{equation}
\label{eq:avg_sum_rate_BS_failure}
\bar{R}_{\mathrm{sum}}^{R}(t)
=
\begin{cases}
\mathbb{E}\!\left[
\displaystyle\sum_{i=1}^{2}\sum_{k\in\mathcal{K}_i}
R_{k,i}(t)
\right], & t<t_f, \\[8pt]
\mathbb{E}\!\left[
\displaystyle\sum_{k=1}^{K}
R_{k,2}(t)
\right], & t\ge t_f,
\end{cases}
\end{equation}
where \(\mathcal{K}_i\) denotes the set of users served by BS-\(b_i\) before failure, and
$R_{k,i}(t)=
a_{\mathrm{RSMA}}^{(i)}(t)\,r_{k,i}^{\mathrm{RSMA}}(t)+\bigl(1-a_{\mathrm{RSMA}}^{(i)}(t)\bigr)\,r_{k,i}^{\mathrm{OMA}}(t)$. For \(t<t_f\), both BSs are active and the total sum rate is the aggregate rate delivered across the two cells. For \(t\ge t_f\), all users are served exclusively by BS-\(b_2\), and the sum rate reflects the post-failure operating point of the surviving BS. The expectation in eq. \eqref{eq:avg_sum_rate_BS_failure} is taken over the spatially correlated MIMO channels, CSI estimation errors, and residual SIC interference effects. In general, the post-failure sum rate is lower than the pre-failure sum rate because all users must be supported by a single BS with limited resources. \vspace{-4mm}
\subsection{System Throughput With Resilience}
To jointly capture spectral efficiency and reliability, the resilient system throughput at time instant \(t\) is defined as
$T_{\mathrm{sys}}^{R}(t)
=
\bigl(1-P_{\mathrm{out}}^{R}(t)\bigr)\,
\bar{R}_{\mathrm{sum}}^{R}(t)$,
where \(P_{\mathrm{out}}^{R}(t)\) denotes the instantaneous outage ratio under the active operating mode and network state. Before failure, both BSs contribute to user service, yielding a higher and more stable throughput region. At \(t=t_f\), the failure of BS-\(b_1\) causes an immediate reduction in available transmission resources and typically produces a sharp throughput drop. For \(t>t_f\), BS-\(b_2\) reconfigures transmission through user consolidation and adaptive RSMA--OMA switching. This enables partial throughput recovery, although the post-failure steady-state throughput is generally lower than the pre-failure level because the surviving BS operates under increased load. Over a horizon of \(T\) time slots, the average resilient throughput is defined as
\begin{equation}
\label{eq:throughput_avg_time_BS_failure}
\bar{T}_{\mathrm{sys}}^{R}
=
\frac{1}{T}
\sum_{t=1}^{T}
T_{\mathrm{sys}}^{R}(t).
\end{equation}

The metric in eq. \eqref{eq:throughput_avg_time_BS_failure} quantifies the long-term ability of the network to sustain useful throughput despite BS failure, channel uncertainty, and receiver impairments.\vspace{-4mm}

\subsection{Discussion of the Resilience Mechanism}

The proposed resilience model reflects a reconfiguration-based recovery process. Before failure, traffic is distributed across two BSs. After the failure of BS-\(b_1\), all affected users are reassigned to BS-\(b_2\), which then re-evaluates the feasibility of RSMA and OMA under the new post-failure channel and load conditions. If RSMA remains feasible and sufficiently reliable, the surviving BS continues transmission in RSMA mode. Otherwise, it activates OMA as a fallback strategy. This adaptive behavior is important because the operating point after failure can differ substantially from the nominal regime. In particular, the surviving BS may face stronger interference coupling, tighter power constraints, and a higher outage risk due to the increased number of served users. Under such conditions, the relative advantage of RSMA may diminish, and the fallback to OMA can improve robustness even if it sacrifices some spectral efficiency. Therefore, the proposed framework does not interpret resilience solely as maintaining the pre-failure access mode. Instead, resilience is defined as the ability to preserve target-rate support and nonzero system throughput through dynamic access-mode adaptation after a BS outage. \vspace{-4mm}

\subsection{Resilience-Aware Throughput Evaluation Algorithm}

The complete resilience-aware throughput evaluation procedure is summarized in Algorithm~\ref{alg:resilience_throughput}.

\begin{algorithm}[t]
\caption{Resilience-Aware Throughput Evaluation Under Adaptive RSMA--OMA}
\label{alg:resilience_throughput}
\footnotesize
\begin{algorithmic}[1]

\STATE \textbf{Input:} $T$, $t_f$, $\{r_{t,k}\}$, $\{P_t^{(1)},P_t^{(2)}\}$, $P_{\mathrm{th}}$, channel statistics, CSI errors, residual SIC interference factors

\STATE \textbf{Output:} $\{T_{\mathrm{sys}}^{R}(t)\}$, $\{\mathcal{R}_{\mathrm{BS}}(t)\}$, $\bar{T}_{\mathrm{sys}}^{R}$

\FOR{$t=1$ to $T$}

\STATE Generate channels, imperfect CSI, and compute effective gains/rates

\IF{$t<t_f$}
\STATE $\mathcal{B}(t)=\{b_1,b_2\}$; determine $\mathcal{K}_1,\mathcal{K}_2$
\FOR{each $b_i\in\mathcal{B}(t)$}
\STATE Compute $P_{\mathrm{RSMA}}^{(i),\mathrm{req}}(t)$ and $P_{\mathrm{out}}^{(i),\mathrm{RSMA}}(t)$; set $a_{\mathrm{RSMA}}^{(i)}(t)=1$ if feasible, else use OMA
\ENDFOR
\STATE Compute $\bar{R}_{\mathrm{sum}}^{R}(t)$
\ELSE
\STATE $\mathcal{B}(t)=\{b_2\}$; reassign all users
\STATE Evaluate RSMA feasibility at $b_2$ and select RSMA/OMA accordingly
\STATE Compute $\{r_k^{(2)}(t)\}$, $\mathcal{R}_{\mathrm{BS}}(t)$, and $\bar{R}_{\mathrm{sum}}^{R}(t)$
\ENDIF

\STATE Compute:
$
P_{\mathrm{out}}^{R}(t),\quad
T_{\mathrm{sys}}^{R}(t)=(1-P_{\mathrm{out}}^{R}(t))\bar{R}_{\mathrm{sum}}^{R}(t)$
\ENDFOR

\STATE Compute:
$
\bar{T}_{\mathrm{sys}}^{R}=\frac{1}{T}\sum_{t=1}^{T}T_{\mathrm{sys}}^{R}(t)$

\end{algorithmic}
\end{algorithm}
\vspace{-4mm}
\subsection{Throughput Recovery Ratio}
To quantify the resilience capability of the proposed adaptive framework under failure conditions, we define the \emph{Throughput Recovery Ratio (TRR)} as the ratio between the steady-state throughput after failure and the steady-state throughput before failure. Mathematically, it is expressed as $\mathrm{TRR}
=\frac{R_{\text{post}}^{\mathrm{ss}}}{R_{\text{pre}}^{\mathrm{ss}}}$, where $R_{\text{pre}}^{\mathrm{ss}}$ and $R_{\text{post}}^{\mathrm{ss}}$ denote the steady-state average sum rates before and after the failure event, respectively.
In the context of the considered RSMA/OMA framework, the steady-state throughput is defined as the long-term time-average (or ensemble average) of the instantaneous achievable sum rate. Specifically, the pre-failure steady-state RSMA throughput is given by
$R_{\text{pre}}^{\mathrm{ss}}
=
\mathbb{E}\left[
\sum_{k=1}^{K} \log_2\bigl(1+\gamma_{p,k}^{\text{pre}}(t)\bigr)
+
\log_2\bigl(1+\gamma_{c}^{\text{pre}}(t)\bigr)
\right]$,
where $\gamma_{p,k}^{\text{pre}}(t)$ and $\gamma_{c}^{\text{pre}}(t)$ represent the private-stream and common-stream SINRs under nominal operating conditions (e.g., accurate CSI and ideal SIC). For the OMA mode, the corresponding pre-failure throughput is expressed as
$R_{\text{pre,OMA}}^{\mathrm{ss}}=
\mathbb{E}\left[
\sum_{k=1}^{K} \tau_k(t)\log_2\left(1+\frac{P_k(t)\|\mathbf{H}_k(t)\|_F^2}{\sigma^2}\right)
\right]$.
After a failure event (e.g., degraded CSI or residual SIC interference), the post-failure steady-state throughput is similarly defined as
$R_{\text{post}}^{\mathrm{ss}}=\mathbb{E}\left[
\sum_{k=1}^{K} \log_2\bigl(1+\gamma_{p,k}^{\text{post}}(t)\bigr)
+
\log_2\bigl(1+\gamma_{c}^{\text{post}}(t)\bigr)
\right]$,
where $\gamma_{p,k}^{\text{post}}(t)$ and $\gamma_{c}^{\text{post}}(t)$ incorporate the effects of $\epsilon^2$ and $\xi$. In particular, the effective channel gains and interference terms are degraded, resulting in lower SINR. The post-failure OMA throughput follows an analogous expression with imperfect CSI affecting the channel gains.
The TRR provides a normalized measure of how effectively a communication scheme can recover its performance after disruption. A value of $\mathrm{TRR} \approx 1$ indicates near-complete recovery, while lower values reflect significant degradation due to failure and system impairments.
In the proposed adaptive RSMA/OMA framework, the system dynamically switches between RSMA and OMA based on instantaneous feasibility conditions, thereby enabling efficient utilization of available resources even under degraded channel states. As a result, the adaptive scheme is expected to achieve a higher TRR compared to fixed RSMA and fixed OMA schemes. Specifically, fixed RSMA suffers from residual interference and imperfect SIC under high impairment levels, leading to reduced post-failure throughput, whereas fixed OMA lacks interference management flexibility \cite{Clerckx2019RSMA}, limiting its recovery performance.
Furthermore, TRR is strongly influenced by channel estimation errors and residual SIC interference factors. As $\epsilon^2$ and $\xi$ increase, both $R_{\text{pre}}^{\mathrm{ss}}$ and $R_{\text{post}}^{\mathrm{ss}}$ degrade; however, degradation is more pronounced in fixed schemes due to their lack of adaptability. In contrast, the proposed adaptive framework maintains a relatively higher TRR by opportunistically selecting the more robust transmission mode, enhancing system reliability and robustness in dynamic wireless environments.

\section{Results and Discussions}
This section presents numerical results to evaluate the performance of the resilience-enabled, degeneracy-aware RSMA framework under imperfect CSI and residual SIC interference. Monte Carlo simulations are conducted to validate the analytical expressions and assess the impact of transmit SNR, SIC interference, CSI uncertainty, and BS failures on sum rate. The resilient framework, with two BSs ($b_1$ and $b_2$), ensures service continuity through user reassignment during failures. The proposed scheme is compared with a non-resilient RSMA benchmark to highlight the performance gains from degeneracy awareness and resilience mechanisms. The specific simulation parameters used in this work are summarized in Table~\ref{tab:sim_params}.
\begin{table}[t]
\centering
\caption{Simulation Parameters}
\label{tab:sim_params}
\begin{tabular}{|c|c|}
\hline
\textbf{Parameter} & \textbf{Value} \\ \hline
Number of Users ($K$) & $\{2,4,6,8,10,12\}$ \\ \hline
CSI Error Variance ($\epsilon^2$) & $0$ to $0.3$ \\ \hline
Residual SIC Interference Factor ($\xi$) & $0$ to $0.3$ \\ \hline
Noise Power ($\sigma^2$) & $1$ \\ \hline
SNR Range & $0$ dB to $30$ dB \\ \hline
Low SNR & $5$ dB \\ \hline
High SNR & $25$ dB \\ \hline
Transmit Power ($P_t$) & $10$ dB \\ \hline
Target Rate ($r_{t,k}$) & $1$ bps/Hz \\ \hline
\end{tabular}
\end{table}

\begin{figure}[t]
    \centering
    \includegraphics[width=\linewidth]{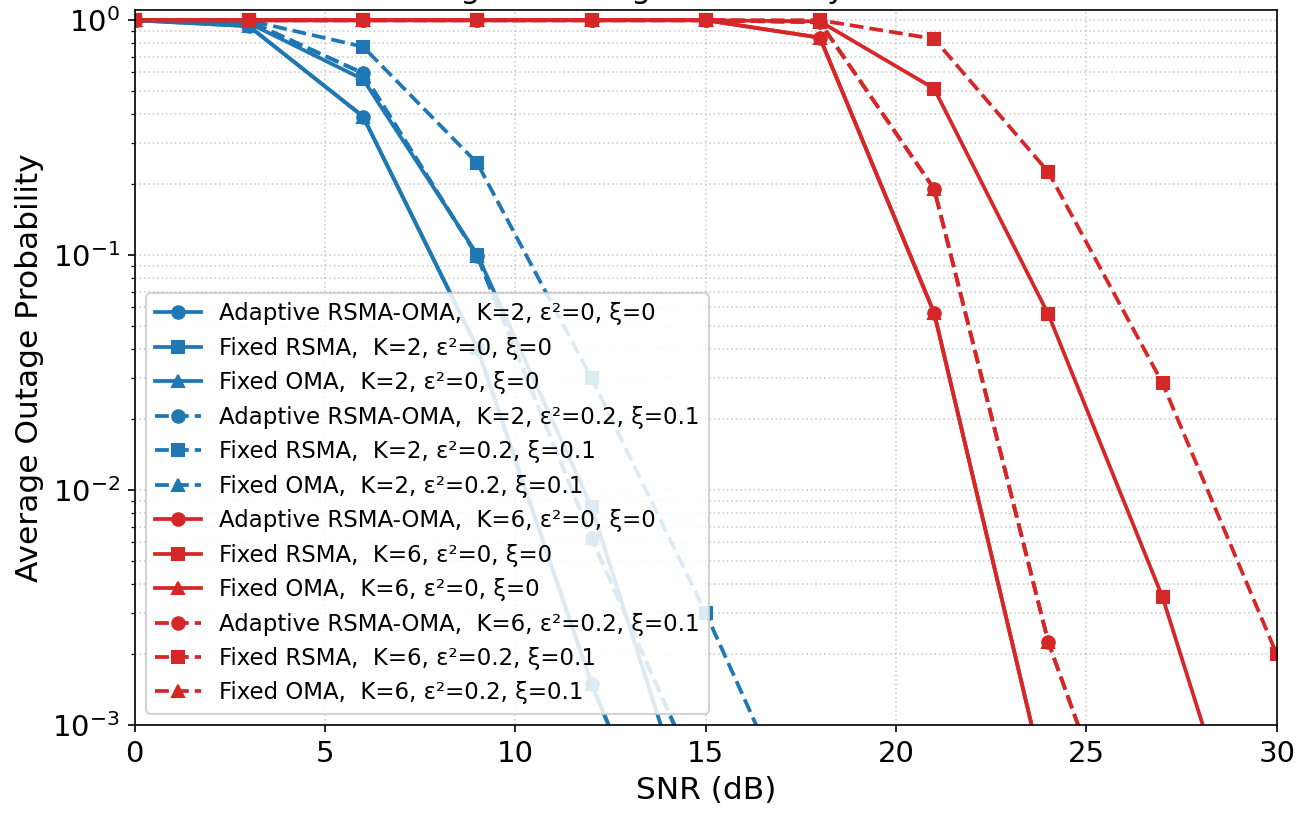}
    \vspace{-3mm}
    \caption{Average outage probability versus SNR for adaptive RSMA-OMA, fixed RSMA, and fixed OMA schemes under different user densities ($K = 2, 6$) and impairment conditions. The simulation considers imperfect CSI with error variance $\epsilon^2 \in \{0, 0.2\}$ and residual SIC interference factor $\xi \in \{0, 0.1\}$. The noise power is $\sigma^2 = 1$, and the target rate for each user is $r_{t,k} = 1$ bps/Hz.}
    \label{Fig.2}
\vspace{-6mm}
\end{figure}

Fig.~\ref{Fig.2} shows the average outage probability versus transmit SNR for adaptive RSMA--OMA, fixed RSMA, and fixed OMA under different user densities ($K=2$ and $K=6$) and impairment levels. In the low-SNR regime ($0$--$5$~dB), the outage probability remains close to 1 for all schemes because the received signal power is insufficient to satisfy the target-rate constraints under strong noise. In this region, the RSMA activation probability is also very small (see Fig.~\ref{Fig.6}), particularly for the higher user-density case ($K=6$), and the adaptive framework therefore operates predominantly in OMA mode. Consequently, the adaptive RSMA--OMA and fixed OMA curves become very close at low SNR values. As SNR increases, the outage probability decreases for all schemes, although the transition occurs at different SNR ranges depending on the user density and impairment conditions. Under ideal CSI and SIC conditions ($\epsilon^2=0$, $\xi=0$), the adaptive RSMA--OMA and fixed RSMA schemes achieve outage reduction at lower SNR values than fixed OMA, because RSMA more effectively manages multiuser interference through common and private stream transmission. Under imperfect CSI and residual SIC interference ($\epsilon^2=0.2$, $\xi=0.1$), all curves shift toward higher SNR values since CSI uncertainty and residual interference increase the effective interference level and the transmit power required to satisfy the QoS constraints derived in Section~III. In this impaired regime, the adaptive framework follows the performance trend of fixed RSMA at moderate and high SNR while retaining the ability to switch to OMA whenever the RSMA feasibility condition in \eqref{eq:28} is not satisfied. At high SNR, the RSMA activation probability approaches 1 (Fig.~\ref{Fig.6}), and the adaptive RSMA--OMA and fixed RSMA curves therefore become nearly overlapping for both $K=2$ and $K=6$. Increasing the number of users from $K=2$ to $K=6$ shifts the outage curves toward higher SNR values due to stronger multiuser interference and increased competition for the available transmit power.\\
\begin{figure}[t]
    \centering
    \includegraphics[width=\linewidth]{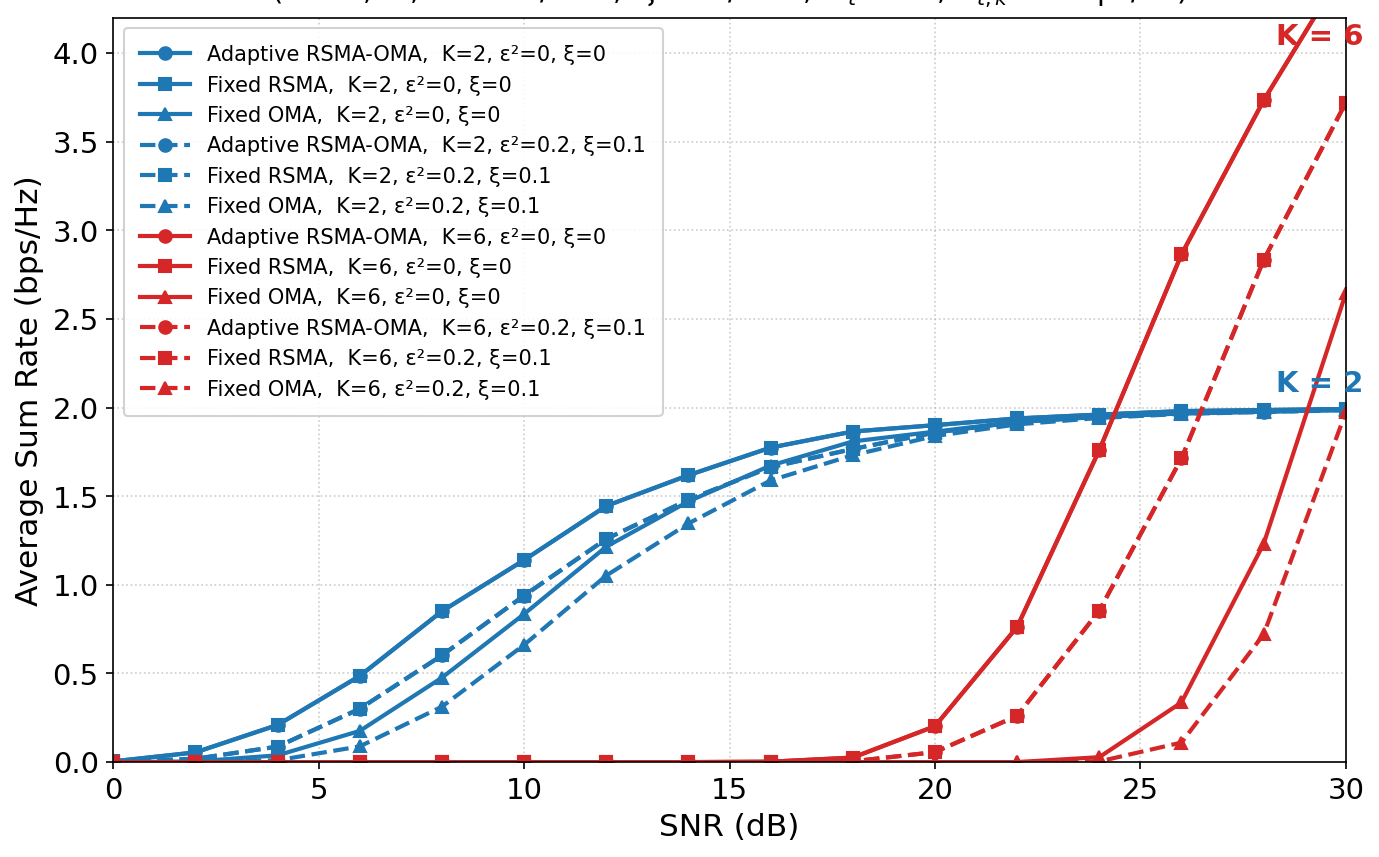}
    \vspace{-3mm}
    \caption{Achievable sum rate versus SNR for adaptive RSMA-OMA, fixed RSMA, and fixed OMA schemes under varying user densities ($K = 2, 6$) and impairment levels. The CSI error variance and residual SIC interference factor are set to $\epsilon^2 \in \{0, 0.2\}$ and $\xi \in \{0, 0.1\}$, respectively.}
    \label{Fig.3}
   \vspace{-7mm}
\end{figure}
Fig.~\ref{Fig.3} illustrates the achievable sum rate versus SNR for adaptive RSMA--OMA, fixed RSMA, and fixed OMA under different user densities ($K=2$ and $K=6$) and impairment conditions. For all schemes, the achievable sum rate increases with SNR because higher transmit power improves the effective SINR of both the common and private streams. The impact of user density is also evident: the $K=6$ case achieves a higher maximum sum rate than $K=2$ due to the increased spatial multiplexing gain obtained from serving more users simultaneously. For $K=2$, the adaptive RSMA--OMA and fixed RSMA curves remain very close across most of the SNR range under both ideal and impaired conditions, indicating that RSMA remains feasible for a large fraction of channel realizations when the number of users is small. The overlap becomes more pronounced at high SNR, where the RSMA activation probability approaches one (Fig.~\ref{Fig.6}), making the adaptive scheme effectively equivalent to fixed RSMA. In contrast, fixed OMA achieves a lower sum rate because orthogonal resource allocation limits simultaneous spatial transmission gains. For the higher user-density case ($K=6$), the differences among the schemes become more visible. Under ideal conditions ($\epsilon^2=0$, $\xi=0$), the adaptive RSMA--OMA and fixed RSMA schemes achieve higher sum rates than fixed OMA, because RSMA more efficiently manages multiuser interference through common and private stream transmission. Under imperfect CSI and residual SIC interference ($\epsilon^2=0.2$, $\xi=0.1$), all curves shift toward higher SNR values due to degraded channel estimation accuracy and increased residual interference. In this regime, the adaptive framework provides comparable or slightly better performance than fixed RSMA over part of the transition region, since it can switch to OMA whenever the RSMA feasibility condition becomes difficult to satisfy. At high SNR, the adaptive and fixed RSMA curves again become close because RSMA becomes consistently feasible and reliable, reducing the need for fallback switching.
\begin{figure}[t]
    \centering    \includegraphics[width=0.98\linewidth]{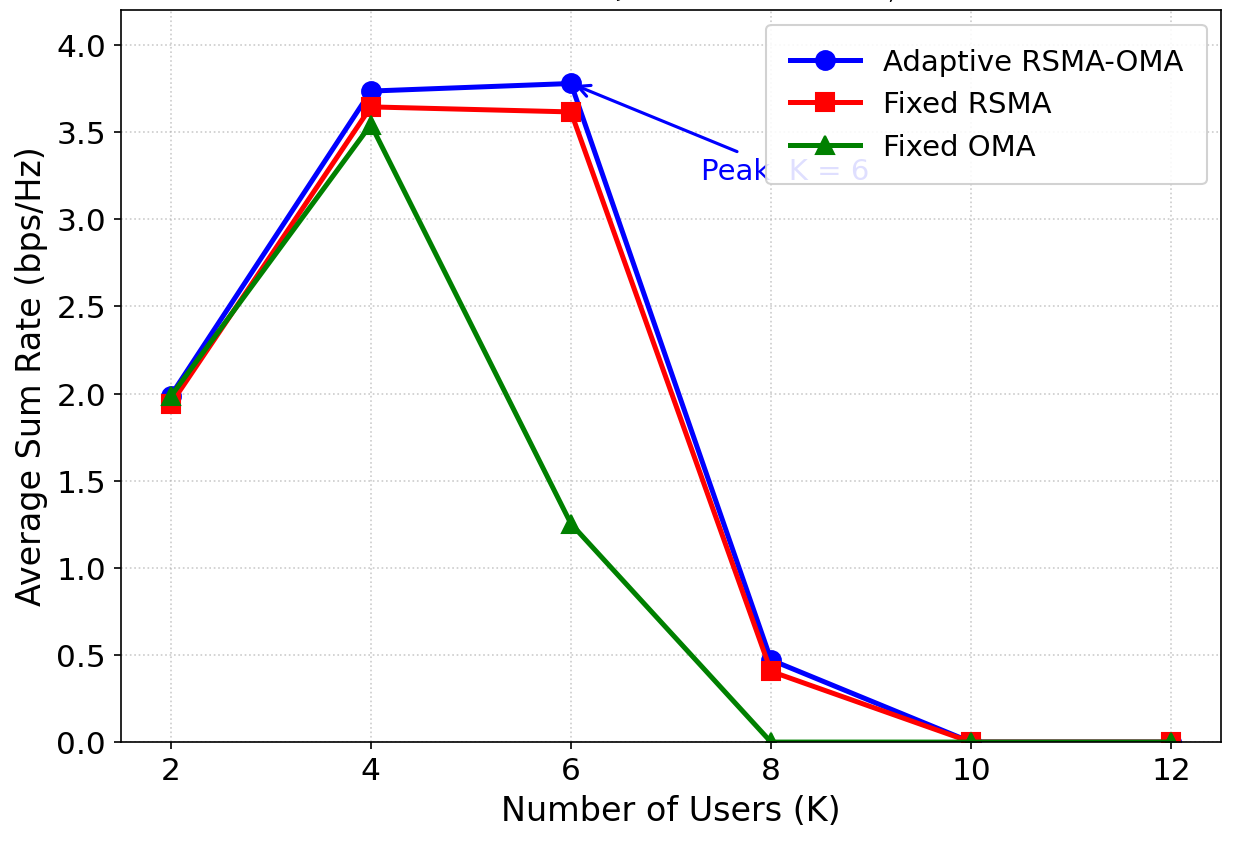}
   \caption{Average sum rate versus number of users ($K$) for adaptive RSMA-OMA, fixed RSMA, and fixed OMA schemes. The simulation is conducted at a fixed SNR (e.g., 15 dB) with noise power $\sigma^2 = 1$, target rate $r_{t,k} = 1$ bps/Hz, and moderate impairment conditions ($\epsilon^2 = 0.2$, $\xi = 0.1$).}
    \label{Fig.4}
  \vspace{-7mm}
\end{figure}

Fig.~\ref{Fig.4} illustrates the average sum rate versus the number of users $K$ for adaptive RSMA--OMA, fixed RSMA, and fixed OMA at a fixed SNR of $15$~dB under moderate impairments ($\epsilon^2=0.2$, $\xi=0.1$). The adaptive RSMA--OMA and fixed RSMA schemes initially benefit from increasing user density because additional users provide higher spatial multiplexing opportunities and improved spectrum utilization. As $K$ increases from $2$ to $6$, the sum rate of these two schemes increases since the multiplexing gain dominates the interference growth in this operating region. In contrast, the fixed OMA scheme exhibits only limited or no improvement with increasing $K$ because orthogonal resource allocation restricts simultaneous transmission efficiency and reduces the available resource fraction per user. The peak around $K=6$ for the adaptive RSMA--OMA and fixed RSMA schemes corresponds to the operating point where multiplexing gain and interference are approximately balanced. Beyond this point, the sum rate decreases due to severe multiuser interference, tighter power-allocation constraints, and increased decoding complexity. In this dense-user regime, RSMA activation probability decreases (Fig.~\ref{Fig.6}) because satisfying the RSMA feasibility conditions becomes increasingly difficult under strong interference coupling. Consequently, the adaptive framework switches more frequently to OMA operation to maintain feasible transmission. Fixed RSMA experiences a more pronounced degradation at large $K$ because it remains constrained to a single interference-management strategy even when residual SIC interference and CSI uncertainty significantly deteriorate the achievable SINR. Fixed OMA maintains lowest overall performance due to strict orthogonalization and reduced spectral efficiency.

\begin{figure}[t]
    \centering
    \includegraphics[width=\linewidth]{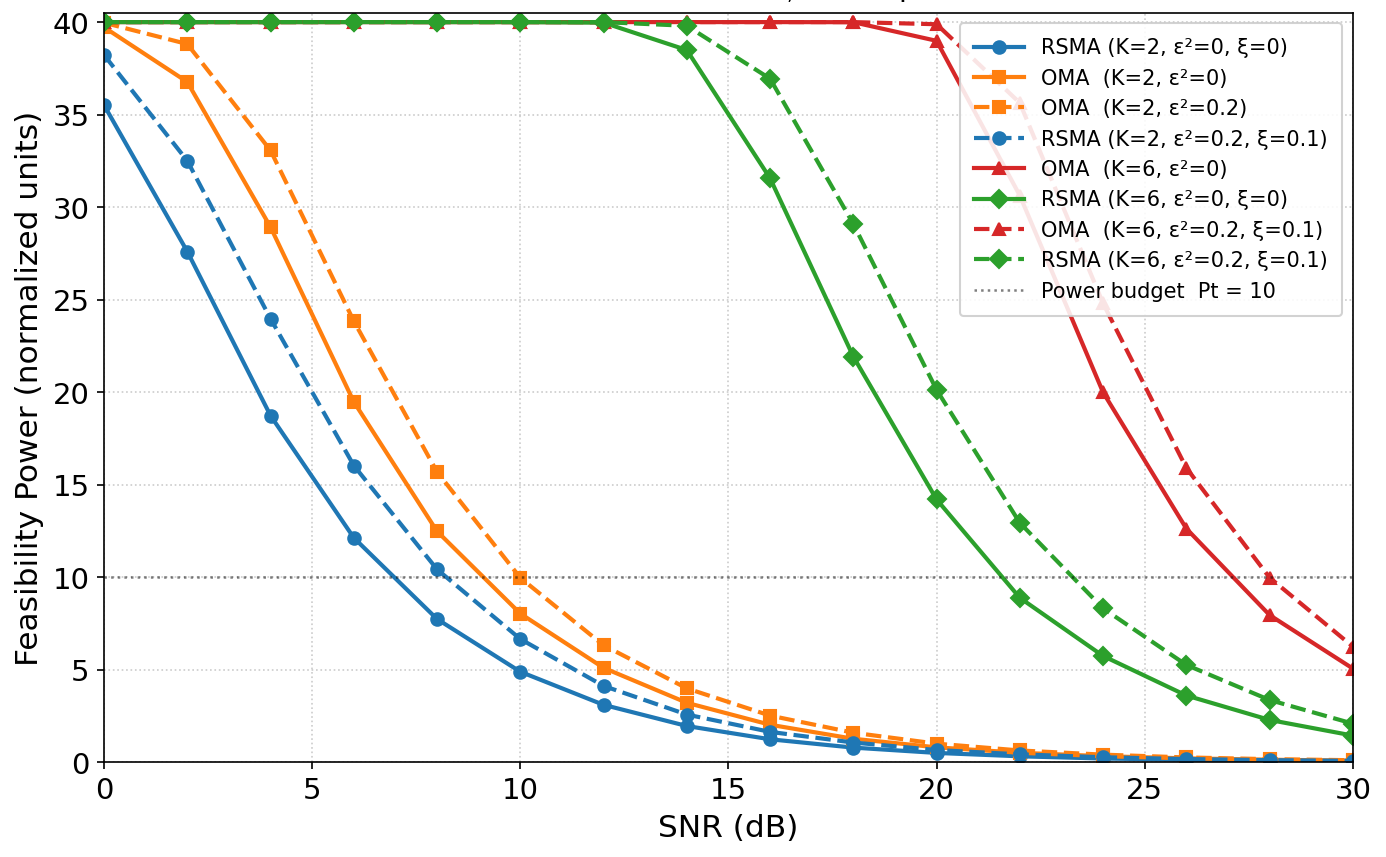}
   \caption{Minimum required (feasibility) transmit power versus SNR for RSMA and OMA schemes under different user densities ($K = 2, 6$) and impairment conditions ($\epsilon^2 \in \{0, 0.2\}$, $\xi \in \{0, 0.1\}$).}
    \label{Fig.5}
  \vspace{-6mm}
\end{figure}

Fig.~\ref{Fig.5} shows the minimum required transmit power versus SNR for
RSMA and OMA under different user densities ($K=2,\,6$) and impairment levels. The required power decreases monotonically with increasing SNR for all schemes. At low SNR, RSMA consistently requires less power than OMA under ideal conditions ($\epsilon^2=0$, $\xi=0$) because common-stream transmission efficiently absorbs a portion of the multiuser interference load. Under impaired conditions ($\epsilon^2=0.2$, $\xi=0.1$), both schemes demand more power;
however, RSMA maintains its advantage through flexible common-private power
splitting. For $K=6$, OMA requires significantly more power than RSMA due to resource partitioning, whereas RSMA scales more efficiently with user density. At high SNR, the feasibility power of both schemes converges toward low values, reducing the performance gap and confirming the complementary roles of RSMA and OMA across different SNR regimes. Importantly, this convergence is also the reason why the adaptive scheme selects RSMA with probability approaching one at high SNR (Fig.~\ref{Fig.6}), the power budget is no longer a binding constraint, RSMA is always feasible, and OMA fallback is never triggered.

\begin{figure}[t]
    \centering
    \includegraphics[width=\linewidth]{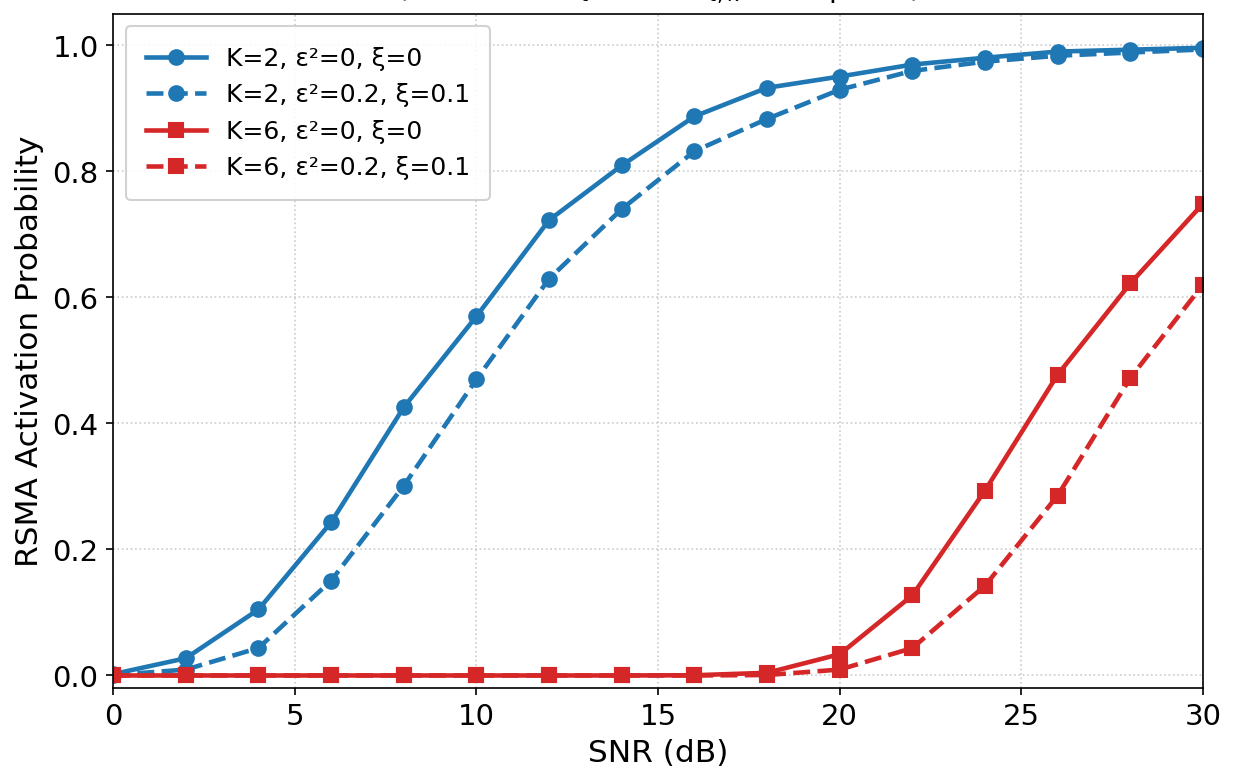}
   \caption{RSMA activation probability versus SNR for the proposed adaptive RSMA-OMA framework under varying user densities ($K = 2, 6$) and impairment conditions. The CSI error variance and residual SIC interference  are $\epsilon^2 \in \{0, 0.2\}$ and $\xi \in \{0, 0.1\}$, respectively.}
    \label{Fig.6}
   \vspace{-6mm}
\end{figure}

\begin{figure}[t]
    \centering
    \includegraphics[width=\linewidth]{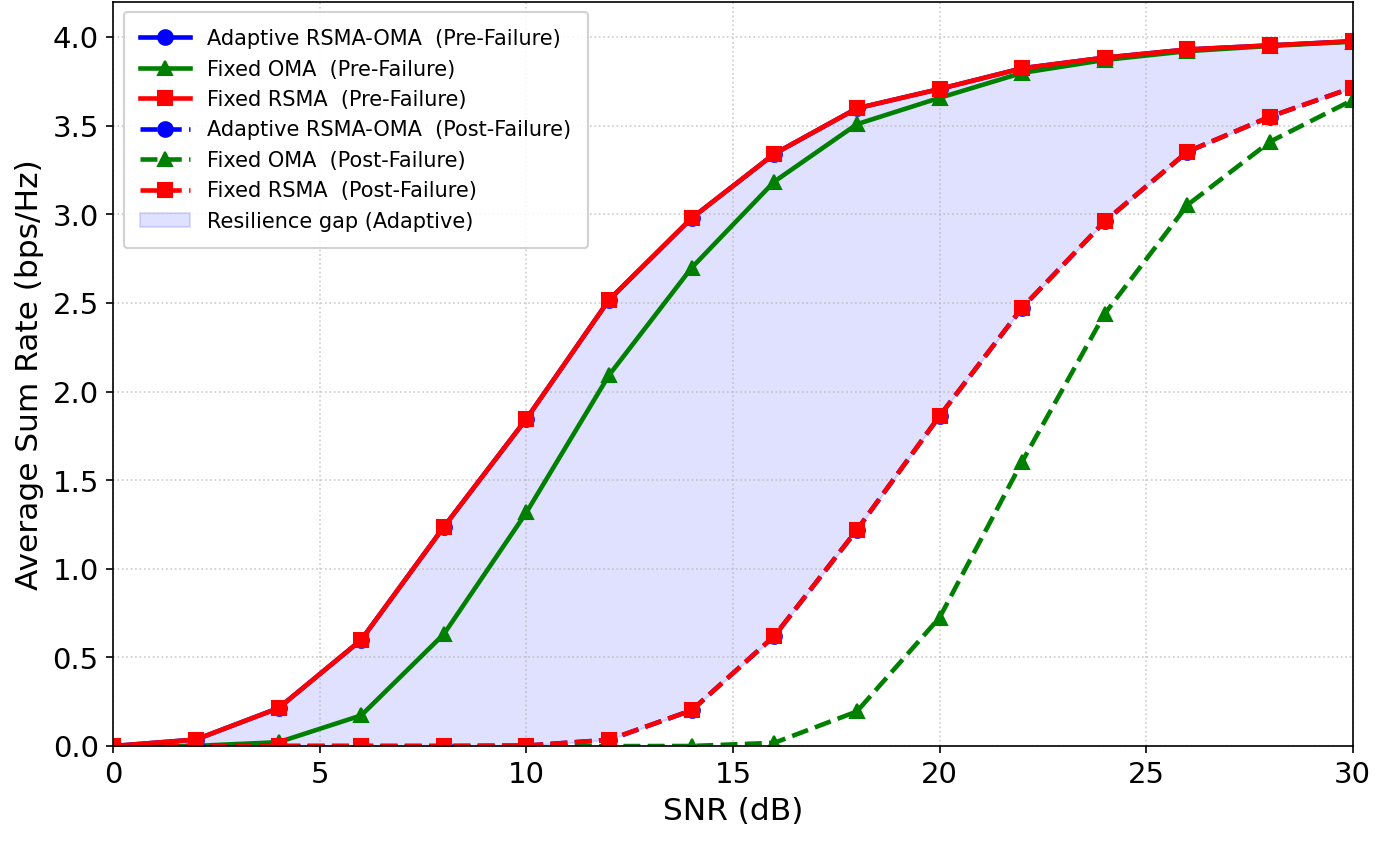}
   \caption{Average sum rate versus SNR under pre-failure and post-failure (resilience) scenarios for adaptive RSMA-OMA, fixed RSMA, and fixed OMA schemes. The simulation considers imperfect CSI error ($\epsilon^2 = 0.2$) and residual SIC interference ($\xi = 0.1$). In the post-failure regime, all users are served by a single surviving BS, leading to increased interference and resource constraints.}
    \label{Fig.7}
    \vspace{-7mm}
\end{figure}

Fig.~\ref{Fig.6} illustrates the RSMA activation probability versus SNR for different user densities and impairment levels. This figure is important for interpreting the adaptive framework, because it directly quantifies how frequently the system operates in RSMA mode instead of OMA mode under varying channel conditions. The activation probability increases monotonically with SNR for all scenarios, and approaches one at sufficiently high SNR. At low SNR, the RSMA activation probability remains close to zero, particularly for the higher user-density case ($K=6$), because the transmit power required to satisfy all target-rate constraints under RSMA exceeds the available power budget $P_t$. Consequently, the adaptive framework operates predominantly in OMA mode in this regime, which explains why the adaptive curves in Figs.~\ref{Fig.2} and~\ref{Fig.3} become close to the fixed OMA curves at low SNR values. As SNR increases, the feasibility conditions become easier to satisfy and the RSMA activation probability gradually increases. Under ideal conditions ($\epsilon^2=0$, $\xi=0$), RSMA becomes feasible at lower SNR values than under impaired conditions ($\epsilon^2=0.2$, $\xi=0.1$), since CSI uncertainty and residual SIC interference increase the effective interference level and tighten the RSMA feasibility constraints. The effect of user density is also clearly visible. Increasing the number of users from $K=2$ to $K=6$ shifts the activation curves toward higher SNR values, because additional users introduce stronger interference coupling and require higher transmit power for simultaneous QoS satisfaction. At high SNR, the activation probability approaches one, implying that the adaptive scheme operates almost entirely in RSMA mode. Consequently, the adaptive curves in Figs.~\ref{Fig.2} and~\ref{Fig.3}, and as we will see next, Fig.~\ref{Fig.7}, become very close to the fixed RSMA curves in the high-SNR region.

Fig.~\ref{Fig.7} shows the average sum rate versus SNR under pre-failure and post-failure conditions for adaptive RSMA--OMA, fixed RSMA, and fixed OMA under imperfect CSI ($\epsilon^2=0.2$) and residual SIC interference ($\xi=0.1$). In the pre-failure regime, both BSs are operational and jointly serve the users. For all schemes, the achievable sum rate increases with SNR because higher transmit power improves the effective SINR and reduces the impact of noise. The pre-failure adaptive RSMA--OMA and fixed RSMA curves remain very close over most of the SNR range, particularly at moderate and high SNR, indicating that RSMA remains feasible for a large fraction of channel realizations before the failure event. At high SNR, the two curves almost overlap because the RSMA activation probability approaches one (Fig.~\ref{Fig.6}), so the adaptive scheme operates predominantly in RSMA mode. Fixed OMA achieves a lower sum rate because orthogonal resource allocation limits simultaneous transmission efficiency. After the failure event, all users are reassigned to the surviving BS $b_2$, which increases the traffic load and interference level experienced by the remaining transmitter. Consequently, the post-failure sum-rate curves shift toward higher SNR values compared with the pre-failure case, reflecting the increased power requirement needed to satisfy the same QoS constraints using only one active BS. This degradation is more visible in the low-to-moderate SNR region, where the surviving BS faces tighter feasibility constraints due to user consolidation. The adaptive RSMA--OMA and fixed RSMA post-failure curves again become close at moderate and high SNR values, because RSMA becomes feasible for most channel realizations in this region. However, the adaptive framework exhibits a smoother performance transition after failure, since it can switch between RSMA and OMA depending on the instantaneous feasibility and outage conditions. The shaded region in Fig.~\ref{Fig.7} highlights the resilience gap between pre-failure and post-failure operation for the adaptive scheme, illustrating the throughput loss caused by BS failure and resource consolidation. 

\begin{figure}[t]
    \centering
    \includegraphics[width=\linewidth]{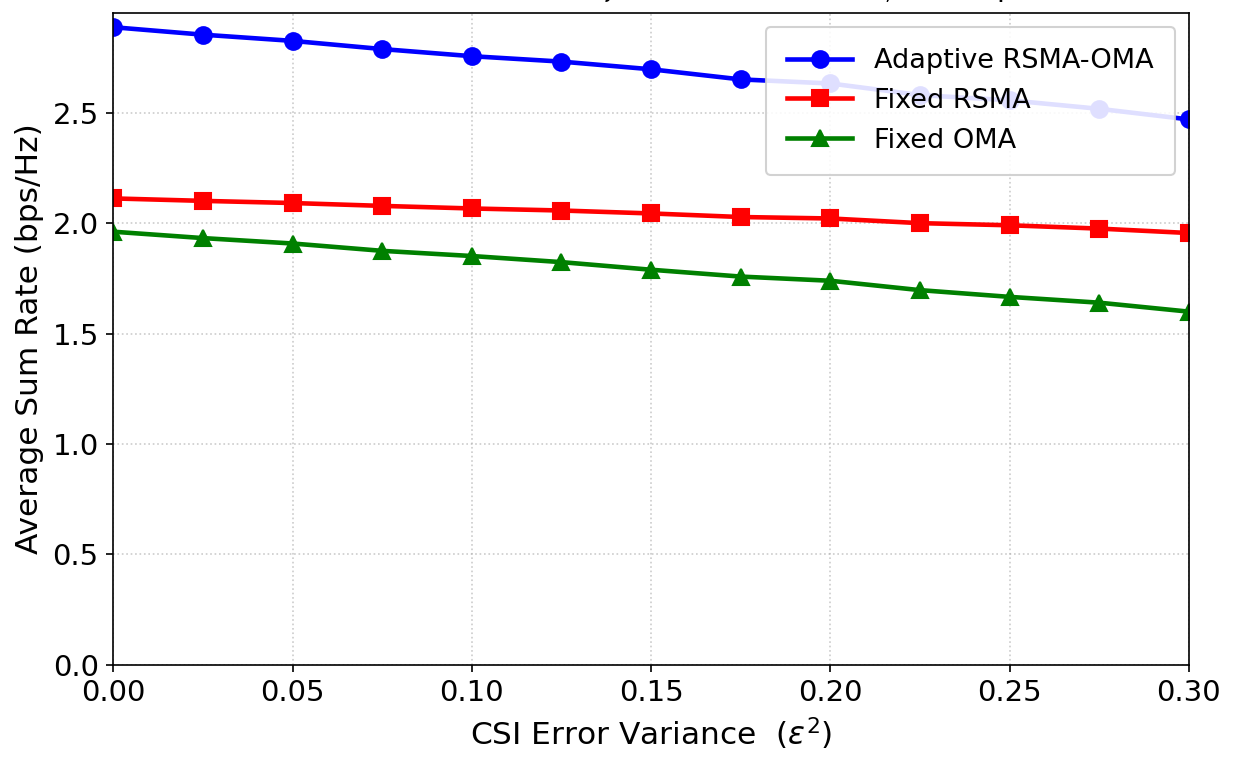}
   \caption{Average sum rate versus CSI error variance $\epsilon^2$ for adaptive RSMA-OMA, fixed RSMA, and fixed OMA schemes at low SNR (5 dB) with $K = 4$ users. The residual SIC interference factor is fixed at $\xi = 0.1$, and the target rate is $r_{t,k} = 1$ bps/Hz.}
    \label{Fig.8}
    \vspace{-7mm}
\end{figure}
Fig.~\ref{Fig.8} illustrates the average sum rate versus CSI error variance $\epsilon^2$ for adaptive RSMA--OMA, fixed RSMA, and fixed OMA at low SNR (5 dB) with $K=4$ users. The average sum rate decreases gradually with increasing $\epsilon^2$ for all schemes because larger CSI uncertainty reduces channel estimation accuracy and degrades the effective SINR. The adaptive RSMA--OMA scheme consistently achieves the highest sum rate across the entire range of CSI error values due to its ability to dynamically select the transmission mode according to the prevailing channel conditions and feasibility constraints. Fixed RSMA maintains higher throughput than fixed OMA throughout the considered CSI error range because RSMA provides more efficient interference management and spectrum utilization through common and private stream transmission. However, both fixed RSMA and adaptive RSMA--OMA experience gradual throughput reduction as $\epsilon^2$ increases, reflecting the sensitivity of RSMA-based transmission to CSI uncertainty. Fixed OMA also exhibits decreasing throughput with increasing $\epsilon^2$, and its performance remains consistently lower because orthogonal resource allocation limits simultaneous spatial transmission efficiency. The relatively smooth degradation of all curves indicates that, at low SNR, the system performance is jointly influenced by both noise and CSI uncertainty rather than by CSI error alone. The adaptive framework preserves the largest overall sum rate because it can flexibly operate between RSMA and OMA depending on the instantaneous feasibility conditions under imperfect CSI. \\
\begin{figure}[t]
    \centering
    \includegraphics[width=\linewidth]{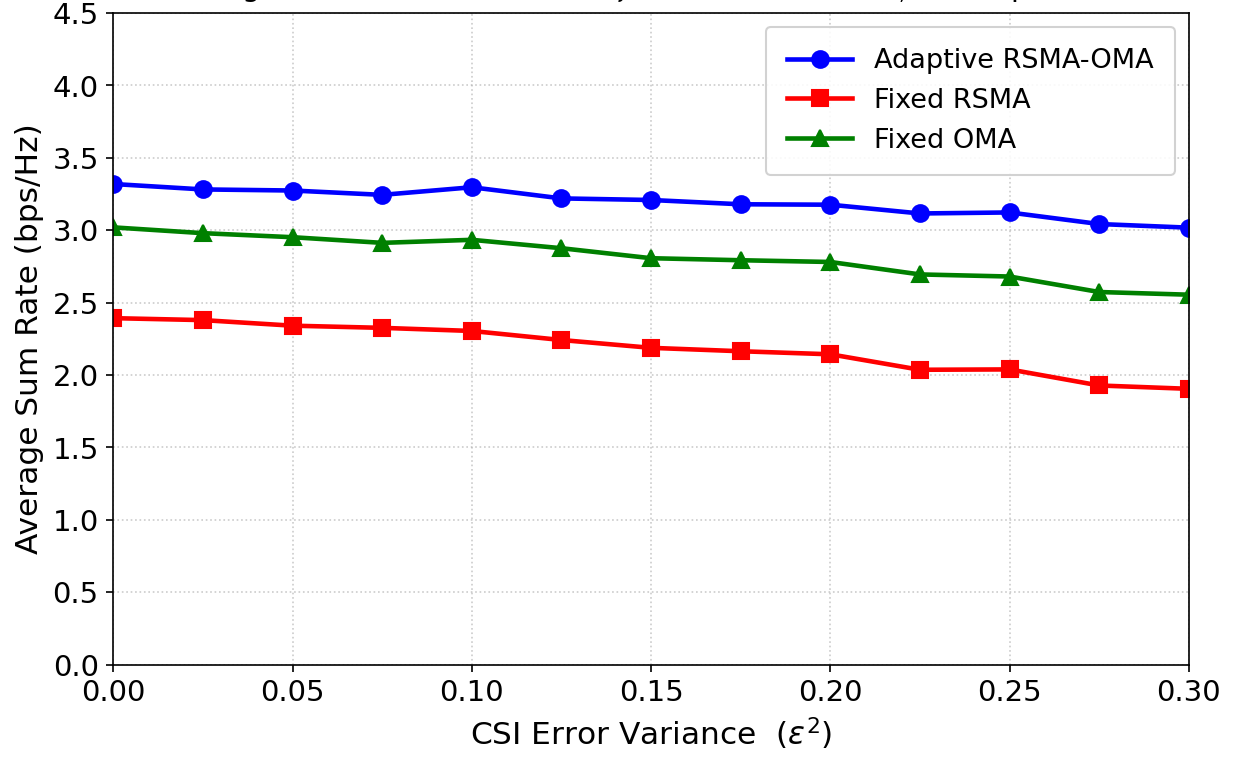}
   \caption{Average sum rate versus CSI error variance $\epsilon^2$ for adaptive RSMA-OMA, fixed RSMA, and fixed OMA schemes at high SNR (25 dB) with $K = 4$ users. The residual SIC interference factor is $\xi = 0.1$.}
    \label{Fig.9}
    \vspace{-6mm}
\end{figure}
Fig.~\ref{Fig.9} illustrates average sum rate versus CSI error variance $\epsilon^2$ for adaptive RSMA--OMA, fixed RSMA, and fixed OMA at high SNR (25 dB) with $K=4$ users. The average sum rate gradually decreases with increasing CSI error variance for all schemes because imperfect channel estimation reduces accuracy of interference management and effective beamforming. At high SNR, the system transitions from a noise-limited regime to an interference- and CSI-limited regime. Adaptive RSMA--OMA scheme consistently achieves the highest sum rate across the entire range of CSI error values due to its ability to dynamically operate between RSMA and OMA according to the instantaneous channel and feasibility conditions. Fixed OMA achieves intermediate performance, while fixed RSMA exhibits the lowest sum rate for the considered parameter setting. This behavior indicates that, under the combined effect of high SNR, imperfect CSI, and residual SIC interference, the fixed RSMA configuration becomes more sensitive to interference mismatch and decoding imperfections. In contrast, the adaptive framework maintains higher throughput by flexibly adapting the transmission strategy under CSI uncertainty. Although all curves decrease with increasing $\epsilon^2$, the degradation remains relatively gradual over the considered range, indicating that the proposed adaptive framework preserves stable operation even under significant channel estimation errors.\\
\begin{figure}[t]
    \centering
    \includegraphics[width=\linewidth]{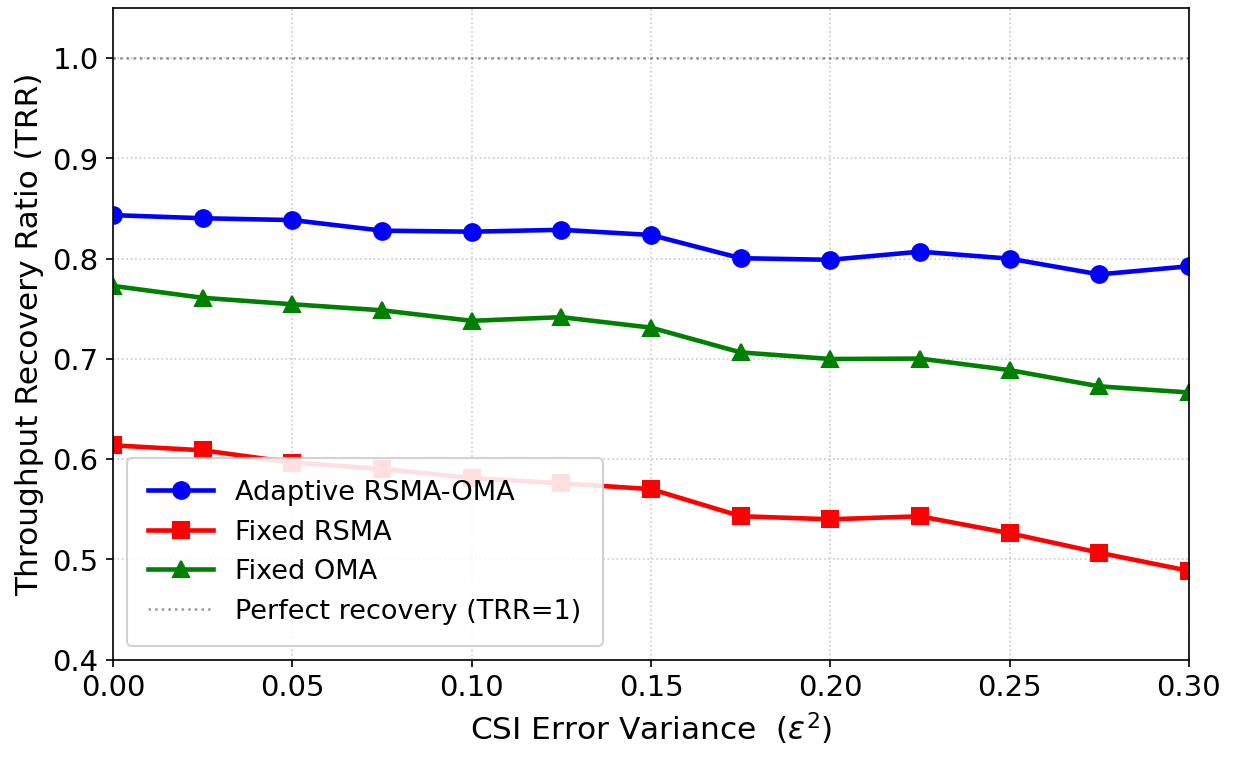}
    \caption{Throughput Recovery Ratio (TRR) versus CSI error variance $\epsilon^2$ for adaptive RSMA-OMA, fixed RSMA, and fixed OMA schemes. The residual SIC interference factor is $\xi = 0.1$, and the number of users is fixed (e.g., $K = 4$).}
    \label{Fig.10}
    \vspace{-7mm}
\end{figure}
Fig. \ref{Fig.10} presents the TRR as a function of the $\epsilon^2$ for the adaptive RSMA--OMA scheme, fixed RSMA, and fixed OMA. The TRR quantifies the ability of the system to retain its pre-impairment performance under increasing channel uncertainty. As observed, the TRR decreases monotonically with increasing $\epsilon^2$ for all schemes, indicating that higher CSI error leads to a significant degradation in the effective system throughput. This behavior is expected, as inaccurate channel estimation deteriorates the SINR by introducing a mismatch in precoding and increasing residual interference, thereby reducing both pre and post-impairment achievable rates. Among the considered schemes, the adaptive RSMA--OMA framework consistently achieves the highest TRR across the entire range of CSI error values. This superior performance arises from its ability to dynamically adapt the transmission strategy based on instantaneous system conditions. In the low CSI error regime, the adaptive scheme predominantly operates in RSMA mode, exploiting its efficient interference management capability to achieve high throughput. As the CSI error increases, the reliability of RSMA degrades due to imperfect channel knowledge and increased residual interference. In such conditions, the adaptive framework switches to OMA, which is less dependent on precise CSI as it avoids interference cancellation and multi-user interference coupling, thereby mitigating severe performance degradation. This dynamic switching mechanism allows the adaptive scheme to maintain a relatively high TRR even under strong channel impairments. The fixed RSMA scheme also shows relatively high TRR at low CSI error levels, as it benefits from accurate channel estimation and effective interference cancellation. However, as $\epsilon^2$ increases, both the adaptive and fixed RSMA schemes exhibit a similar degradation trend, resulting in an approximately constant performance gap between them. This is due to the sensitivity of RSMA to CSI inaccuracies and residual SIC interference $(\xi = 0.1)$, which significantly affect both the common and private stream decoding processes. Consequently, the inability of fixed RSMA to adapt its transmission strategy leads to a reduced capability to recover throughput under severe impairments. In contrast, the fixed OMA scheme exhibits the lowest TRR across all values of $\epsilon^2$. As the CSI error increases, the degradation in channel gain further reduces its achievable rate, leading to a more pronounced decline in TRR.
\section{Conclusion}
This paper introduces a novel degeneracy-aware power control framework for RSMA in MIMO systems under the presence of imperfect CSI and residual SIC interference. By dynamically adjusting power allocation and incorporating an adaptive fallback mechanism to OMA, the proposed framework ensures system feasibility and resilience in challenging communication environments. Extensive simulation results validate that the framework significantly improves power efficiency, reduces outage probability, and enhances system robustness, making RSMA a viable solution for real-world wireless networks with imperfect CSI and SIC. The proposed approach represents a practical and effective enhancement to RSMA performance in realistic, interference-limited communication scenarios.
\vspace{-4mm}
\appendices
\section{Nonconvexity of the RSMA Feasibility Problem}

The nonconvexity of the RSMA feasibility problem in eq. \eqref{eq:fp1}--\eqref{eq:fp5} primarily stems from the SINR expressions associated with both the private and common streams. Consider the private-stream rate constraint, i.e., $c_k(t)+\log_2\!\bigl(1+\gamma_{p,k}(t)\bigr)\ge r_{t,k}$.
This constraint can be equivalently expressed as
\[
\begin{aligned}
P_k(t)g_{k,k}(t) 
&\ge \Gamma_{p,k}(t)\Bigg(
\sum_{j\neq k} P_j(t)g_{k,j}(t) \\
&\quad + \xi_k(t)P_c(t)g_{c,k}(t) + \sigma^2
\Bigg)
\end{aligned}
\]
where $\Gamma_{p,k}(t)=2^{r_{t,k}-c_k(t)}-1$, which depends nonlinearly on $c_k(t)$, while RHS introduces coupling among multiple power variables, resulting in bilinear and fractional terms. Furthermore, the common-rate constraint $\sum_{k}c_k(t)\le r_c(t)$ is also nonconvex, since $r_c(t)=\min_k \log_2\!\bigl(1+\gamma_{c,k}(t)\bigr)$ involves the minimum of nonconcave functions. As a consequence, the feasible set is not convex in the joint variable space $\bigl(P_c(t),\{P_k(t)\},\{c_k(t)\}\bigr)$. This motivates the adoption of the proposed feasibility-driven allocation framework and adaptive RSMA--OMA switching strategy, rather than relying on direct convex optimization approaches.\vspace{-4mm}
\section{Convexity of the OMA Fallback Problem}

Consider the OMA feasibility problem. The user-$k$ rate constraint is $
\tau_k(t)\log_2\!\left(1 + \frac{P_k^{\mathrm{OMA}}(t)\|\mathbf{H}_k(t)\|_F^2}{\sigma^2}\right) \ge r_{t,k}$. \vspace{-4mm}

\subsection{Case 1: Fixed Resource Fractions}
If $\tau_k(t)$ is fixed, the rate constraint becomes
$P_k^{\mathrm{OMA}}(t) \ge \frac{\sigma^2}{\|\mathbf{H}_k(t)\|_F^2}\left(2^{r_{t,k}/\tau_k(t)} - 1\right)$
which is affine in $P_k^{\mathrm{OMA}}(t)$. Since the objective function is linear and all constraints are affine, the problem is convex. \vspace{-4mm}
\subsection{Case 2: Joint Optimization of Power and Resource Fractions}

If both $\tau_k(t)$ and $P_k^{\mathrm{OMA}}(t)$ are optimization variables, define $x_k(t) = \frac{P_k^{\mathrm{OMA}}(t)}{\tau_k(t)}$. The rate expression becomes:
\begin{equation*}
\tau_k(t)\log_2\left(1 + \frac{x_k(t)\|\mathbf{H}_k(t)\|_F^2}{\sigma^2}\right),
\end{equation*}
which is concave in $(\tau_k(t), x_k(t))$. Therefore, the feasible set is convex after standard perspective reformulation. Unlike the RSMA feasibility problem, the OMA fallback problem admits a convex formulation under standard assumptions.\vspace{-4mm}

\section{Fixed-Point Structure of the RSMA Power Allocation}

The RSMA power-allocation equations in \eqref{eq:25} and \eqref{eq:27} form a coupled system where each private-stream power $P_k^\star(t)$ depends on the other private powers and the common-stream power $P_c^\star(t)$, while $P_c^\star(t)$ depends on all private powers. For fixed $\{c_k^\star(t)\}$, define $\mathbf{p}(t) = \bigl[P_1(t), \dots, P_K(t)\bigr]^T$. The private-stream power equations become $
\mathbf{p}(t) = \mathbf{F}\bigl(\mathbf{p}(t), P_c(t)\bigr)$, where $\mathbf{F}(\cdot)$ is a mapping involving target SINRs, effective channel gains, residual SIC interference, and noise powers. The existence of a feasible RSMA power allocation corresponds to the existence of a nonnegative fixed point of this mapping that also satisfies the common-stream decoding constraints. While closed-form solutions are generally unavailable due to the coupling between streams, the structure suggests iterative methods for updating $\{P_k(t)\}$ and $P_c(t)$, which converge under suitable conditions. This fixed-point structure justifies the use of the minimum required power $P_{\mathrm{RSMA}}^{\mathrm{req}}(t)$ for RSMA feasibility rather than solving a globally convex problem.\vspace{-4mm}

\section{Remarks on Computational Complexity}

RSMA feasibility problem is nonconvex due to the coupled common-rate, private-rate, and interference terms, leading to high computational complexity, especially for large user numbers or rapidly changing channels. The main challenges are: 1) coupled SINR expressions; 2) dependence of private-stream SINR targets on common-rate allocation $\{c_k(t)\}$; and 3) the worst-user common-rate constraint introducing a min-operator. Hence, the problem does not admit a simple closed-form global solution, motivating the use of structured feasibility checks and adaptive mode switching. OMA fallback problem is simpler and can be solved efficiently using convex optimization or closed-form power updates when resource fractions are fixed.
\bibliographystyle{IEEEtran}
\bibliography{references}
\end{document}